\documentclass[12pt]{article}
\usepackage{amsmath}
\usepackage{amsfonts}
\usepackage{graphicx}
\usepackage{color}

\newcommand{\bmat}{\left(\begin{array}}
\newcommand{\emat}{\end{array}\right)}

\def\yzero{\smash{\hbox{$y\kern-4pt\raise1pt\hbox{${}^\circ$}$}}}

\def\beq{\begin{equation}}
\def\eeq{\end{equation}}
\def\beqa{\begin{eqnarray}}
\def\eeqa{\end{eqnarray}}

\def\-{\hphantom{-}}
\def\ov{\overline}
\def\s2{\frac{1}{\sqrt2}}

\def\beq{\begin{equation}}
\def\eeq{\end{equation}}
\def\beqa{\begin{eqnarray}}
\def\eeqa{\end{eqnarray}}
\def\tr{{\rm tr \,}}

\def\IT{\mathbb{T}}

\def\cn{{\cal N}}

\def\Dsl{\,\raise.15ex\hbox{/}\mkern-13.5mu D} 

\def\IC{\bf C}
\def\IS{\bf S}
\def\IR{\bf R}
\def\IZ{\bf Z}

\newcommand{\drawsquare}[2]{\hbox{%
\rule{#2pt}{#1pt}\hskip-#2pt
\rule{#1pt}{#2pt}\hskip-#1pt
\rule[#1pt]{#1pt}{#2pt}}\rule[#1pt]{#2pt}{#2pt}\hskip-#2pt
\rule{#2pt}{#1pt}}

\newcommand{\fund}{\raisebox{-.5pt}{\drawsquare{6.5}{0.4}}}
\newcommand{\Yasymm}{\raisebox{-3.5pt}{\drawsquare{6.5}{0.4}}\hskip-6.9pt%
        \raisebox{3pt}{\drawsquare{6.5}{0.4}}}
\newcommand{\antifund}{\overline{\fund}}
\newcommand{\bYasymm}{\overline{\Yasymm}}

\topmargin
-1.5cm
\textwidth
15.5cm
\textheight
23.5cm
\oddsidemargin
0.7cm
\evensidemargin
1.2cm

\begin{document}

\makeatletter
\@addtoreset{equation}{section}
\makeatother
\renewcommand{\theequation}{\thesection.\arabic{equation}}
\pagestyle{empty}
\rightline{ IFT-UAM/CSIC-07-56}
\rightline{ CERN-PH-TH/2007-216}
\vspace{0.1cm}
\begin{center}
\LARGE{\bf Non-perturbative superpotentials \\
 across lines of marginal stability  \\[12mm]}
\large{I. Garc\'{\i}a-Etxebarria, A.M. Uranga\\[3mm]}
\footnotesize{PH-TH Division, CERN 
CH-1211 Geneva 23, Switzerland\\
 and \\
Instituto de F\'{\i}sica Te\'orica UAM/CSIC,\\[-0.3em]
Universidad Aut\'onoma de Madrid C-XVI, 
Cantoblanco, 28049 Madrid, Spain \\[2mm] }
\small{\bf Abstract} \\[5mm]
\end{center}
\begin{center}
\begin{minipage}[h]{16.0cm}

  We discuss the behaviour of non-perturbative superpotentials in 4d
  $\cn=1$ type II compactifications (and orientifolds thereof) near
  lines of marginal stability, where the spectrum of contributing BPS
  D-brane instantons changes discontinuously. The superpotential is
  nevertheless continuous, in agreement with its holomorphic
  dependence on the closed string moduli. The microscopic mechanism
  ensuring this continuity involves novel contributions to the
  superpotential: As an instanton becomes unstable against decay to
  several instantons, the latter provide a multi-instanton
  contribution which reconstructs that of the single-instanton before
  decay. The process can be understood as a non-perturbative lifting
  of additional fermion zero modes of an instanton by interactions
  induced by other instantons. These effects provide mechanisms via
  which instantons with $U(1)$ symmetry can contribute to the
  superpotential. We provide explicit examples of these effects for
  non-gauge D-brane instantons, and for D-brane gauge instantons
  (where the motions in moduli space can be interpreted as Higgsing,
  or Seiberg dualities).

\end{minipage}
\end{center}
\newpage
\setcounter{page}{1}
\pagestyle{plain}
\renewcommand{\thefootnote}{\arabic{footnote}}
\setcounter{footnote}{0}

\section{Introduction}

Non-perturbative effects in string theory are a key ingredient in the
proper understanding of the theory, and in particular of the dynamics
of compactifications to four dimensions. Already in the early times
non-perturbative effects (in the form of strongly coupled field theory
sectors) were considered to underlie moduli stabilization and
supersymmetry breaking \cite{Dine:1985rz,Derendinger:1985kk}. The
formal developments on euclidean brane instanton effects (see
e.g. \cite{Becker:1995kb,Witten:1996bn, Harvey:1999as,Witten:1999eg}),
in particular D-brane instantons in type II compactifications (or
F/M-theory duals) have led to a variety of (in some cases very
explicit) applications to e.g. moduli stabilization
\cite{Kachru:2003aw,Denef:2004dm,Denef:2005mm} and the generation of
perturbatively forbidden couplings
\cite{Blumenhagen:2006xt,Ibanez:2006da} (see also
\cite{Haack:2006cy,Florea:2006si,Ibanez:2007rs} and
\cite{Blumenhagen:2007zk} for related applications). D-brane
instantons in local D-brane models have also been explored, recovering
field theory gauge instanton effects
\cite{Billo:2002hm,Akerblom:2006hx,Bianchi:2007fx,Argurio:2007vqa,Bianchi:2007wy},
and with realizations of old and new models of supersymmetry breaking
\cite{Argurio:2007qk,Aharony:2007db,Aganagic:2007py}. Other formal
aspects of D-brane instantons have been recently discussed in
e.g. \cite{Billo:2007py,Akerblom:2007uc}.

In this paper we discuss an interesting formal aspect of
non-perturbative superpotentials generated by instantons in string
theory. In 4d ${\cal N}=1$ supersymmetric compactifications of string
theory, non-perturbative contributions to the superpotential arise
from brane instantons with two fermion zero modes, which are saturated
by the $d^2\theta$ superspace integration. These must necessarily be
1/2-BPS branes, so that the fermion zero modes are given by the
Goldstinos of the two broken supersymmetries.

Hence, the non-perturbative superpotential depends on the precise list
of BPS branes (satisfying certain additional constraints, like the
absence of extra fermion zero modes) at a given point in moduli
space. Now it is a well-known fact that the spectrum of BPS branes can
jump discontinuously across lines of marginal stability
\cite{Douglas:2000qw,Douglas:2000ah,Douglas:2000gi,Denef:2000nb,Denef:2001xn,Aspinwall:2004jr}. Namely,
in type IIA compactifications the spectrum of supersymmetric D2-brane
instantons may jump as one moves in complex structure moduli space
(with the geometric interpretation that a supersymmetric 3-cycle may
split in two independent supersymmetric 3-cycles when the complex
structure is changed); similarly for D-brane instantons in type IIB
compactifications as one moves in Kahler moduli space.

It is therefore a natural question whether the non-perturbative
superpotential is continuous across these lines of marginal
stability. This is expected, given that superpotentials are protected
quantities. In fact, an abrupt change in the superpotential would
correspond to a non-holomorphic dependence on the moduli (since
marginal stability walls are typically of codimension one), which is
not compatible with supersymmetry. It turns out that the microscopic
explanation of the continuity of the non-perturbative superpotential
is related to a wealth of previously unnoticed surprises in D-brane
instanton physics. We devote the present paper to uncovering them in a
few illustrative examples, leaving a systematic discussion for future
work.

The first interesting novelty is that multi-instanton processes can
contribute to the non-perturbative superpotential. Consider an
instanton A that contributes to the non-perturbative superpotential,
and which reaches a line of marginal stability where it splits into
two instantons B and C. Although the instantons B and C do not in
general contribute to the non-perturbative superpotential by
themselves, the 2-instanton process involving B and C simultaneously
does lead to a contribution to the superpotential.  A key ingredient
is that extra zero modes of the two individual instantons are
saturated against each other, in such a way that only two fermion zero
modes are left over for the combined system, see Figure
\ref{twoinstanton}.  Although multi-instanton processes have been
extensively studied for $\cn=2$ and $\cn=4$ supersymmetric gauge
theories (see \cite{Dorey:2002ik} for a review), the possibility to
have them generate non-perturbative superpotentials in $\cn=1$
theories has not been considered in the past.  Moreover, our result
implies that the usual strategy to compute the non-perturbative
superpotential by summing all contributions from suitable BPS
instantons may miss important contributions, due to multi-instanton
processes. We present several explicit examples of this phenomenon,
for D-brane instantons with or without interpretation as gauge theory
instantons.

A second interesting novelty arises from regarding the above
2-instanton process as a non-perturbative lifting of fermion zero
modes. In considering the effective 4d interaction generated by say
the instanton B, one needs to consider the possible interaction terms
which may lift fermion zero modes (at the Gaussian level). The above
mechanism corresponds to a non-perturbative contribution to the
interactions of the fermions zero modes of the instanton B induced by
the instanton C, so that the former can contribute to the
superpotential. This will be more explicitly discussed in several
examples.

A final interesting surprise is related to non-gauge D-brane
instantons (these are standard D-brane instantons, but refer to them
as non-gauge, or sometimes exotic, to distinguish them from D-brane
instantons with gauge field theory interpretation). It is usually
considered that, for a D-brane instanton in a perturbative type II
model to contribute to the superpotential, it must have Chan-Paton
symmetry $O(1)$, so that the orientifold projection eliminates some of
the universal fermion zero modes (arising from an accidental $\cn=2$
supersymmetry in the relevant open string sector). We however present
several examples where instantons with $U(1)$ symmetries contribute to
the superpotential, with the extra zero modes being saturated by
interactions in the instanton world-volume effective action.

As a last remark, the continuity of the non-perturbative
superpotential, combined with string duality will lead to new
interesting properties of non-perturbative superpotentials in F-theory
compactifications across certain topology changing phase transitions,
as we discuss in Section~\ref{sec:F-theory}.

Before entering the discussion, we would like to present the problem
of the continuity of the non-perturbative superpotential in terms more
familiar from the model building viewpoint, in the context of the
recent approaches to use non-perturbative superpotentials to stabilize
Kahler moduli in type IIB compactifications \cite{Kachru:2003aw}. For
concreteness, consider a compactification with a gauge sector arising
from stack of D7-branes. In general, such configuration of branes is
supersymmetric at a point (or locus) $P$ in Kahler moduli space. At
other points or loci $Q$ in moduli space, the D7-branes have
misaligned BPS phases and recombine to form bound states, which
correspond to BPS branes at point $Q$. The field theory interpretation
is that Kahler moduli couple as Fayet-Iliopoulos terms to the
D-branes, which trigger processes of Higgsing/unHiggsing in the gauge
theory. In any event, the gauge sector arising from the D7-branes is
different at the points $P$, $Q$. Consider that the gauge sector at
$P$ develops a non-perturbative superpotential for the Kahler moduli,
such that the resulting scalar potential stabilizes the moduli at the
point $Q$. If the non-perturbative superpotential would not be
continuous across the line of marginal stability, we would find
ourselves in the paradoxical situation that the minimum lies at a
point where the original potential is no longer valid. Needless to
say, such behavior would enormously complicate the problem of moduli
stabilization. Happily, superpotentials are far better behaved
quantities, which can be used universally all over moduli space.

In this paper we focus on non-perturbative superpotentials. On general
grounds we expect that other quantities, such as higher derivative
F-terms, arising from BPS instantons with additional fermion zero
modes, are also continuous all over moduli space.  We leave a
systematic understanding for future work, and will be happy to
constrain ourselves to the discussion of the continuity of the
superpotential in a series of illustrative examples.

The paper is organized as follows. In Section~\ref{sec:background} we
discuss some relevant background material on instantons, both gauge
and non-gauge, and we introduce the geometric backgrounds we will
consider. In Section~\ref{sec:non-gauge} we discuss the continuity of
the superpotential for non-gauge instantons, explaining the role of
multi-instanton processes. In Section~\ref{gaugeinst} we go on to
discuss continuity and multi-instanton effects for gauge instantons in
string theory. In particular we describe the continuity of the
superpotential under Seiberg duality. In
Section~\ref{sec:exotic-to-gauge} we study motions in moduli space
that convert gauge instantons into non-gauge instantons, and vice
versa. In Section~\ref{sec:F-theory} we describe the dual realization
of the processes we study in F and M
theory. Section~\ref{sec:conclusions} contains our conclusions, and
finally Appendix~\ref{nosplit} discusses some exotic geometric
processes that evade the assumptions in this paper, and might lead to
discontinuities in the superpotential.

\section{Some background material}
\label{sec:background}

\subsection{Instanton effects}
\label{introinst}

In dealing with euclidean brane instantons in string theory
compactifications, it is convenient to make some general
classifications and distinctions, which are useful for future
reference. For concreteness we focus on D-brane instantons, although
the effects can arise from other brane instantons in dual pictures (a
prototypical example are e.g. euclidean D3-brane instantons in type
IIB on CY-threefolds described as M5-brane instantons in M-theory on
CY-fourfolds \cite{Witten:1996bn}).

A first class of D-brane instantons corresponds to those whose
internal structure is exactly the same as some of the 4d space filling
branes in the background. Namely, in geometric setups, Euclidean
D$p$-branes wrapping the same $(p+1)$-cycle (and carrying the same
world-volume gauge bundle) as some D$(p+4)$-branes in the background
configuration (in more abstract CFT terms, they should be described by
the same boundary state of the internal CFT). Such D-brane instantons
correspond to gauge instantons on the corresponding 4d gauge sector,
and thus reproduce non-perturbative effects arising from strong gauge
dynamics.

\subsubsection{Superpotentials from gauge D-brane instantons}
\label{supogauge}

A prototypical case, which will appear in our examples, is the
generation of the Affleck-Dine-Seiberg superpotential
\begin{eqnarray}
W\, =\,
(N_c-N_f)\, \left(\, \frac{\Lambda^{3N_c-N_f}}{\det M} \,
\right)^{\frac{1}{N_c-N_f}}
\end{eqnarray}
on a set of 4d space filling branes whose low-energy dynamics
corresponds to $SU(N_c)$ SQCD with $N_f$ flavours (with dynamical
scale $\Lambda$; here $M$ denotes the meson fields). For $N_f=N_c-1$
this arises from classical 4d instanton field configurations, and has
been recovered from D-brane instantons in several instances with
different levels of detail
\cite{Acharya:2000gb,Bershadsky:1996gx,Florea:2006si,Akerblom:2006hx}. For
other values of $N_f<N_c$, it does not arise from classical 4d field
configurations, and is obtained indirectly. Alternatively, it can be
obtained by considering the theory on $\IS^1$, where there exist
suitable 3d classical field configurations (sometimes denoted
calorons) leading to a 3d superpotential, which can be argued to
survive in the 4d decompactification limit (with a microscopic
description in terms of putative objects denoted ``fractional
instantons'' or ``merons''). The latter description fits perfectly
with the string theory realization. Indeed, the computation of
e.g. the euclidean D3-brane instanton superpotential in type IIB
configurations with gauge sectors on D7-branes on 4-cycles is usually
described by invoking compactification on a circle in order to use an
M-theory dual. Upon compactification, one may use T-duality, leading
to a picture where instantons are D-branes stretched along the circle
direction, and gauge D-branes are pointlike on it. In this picture the
superpotential is generated by ``fractional'' D-branes, which are
suspended between the gauge D-branes and thus stretch only a fraction
of the period along the circle direction
\cite{Brodie:1998bv}. Equivalently, in the dual M-theory picture, the
gauge D7-branes turn into degenerations of the elliptic fibration,
such that the fiber over the 4-cycle on the base is a sausage of
2-spheres. The ADS superpotential is generated by M5-branes which wrap
the 4-cycle times a 2-sphere (leading to ``fractional'' objects,
in the sense that the standard 4d gauge instanton corresponds to an
M5-brane wrapping the whole fiber) \cite{Bershadsky:1996gx}. We will
often abuse language and regard the 4d ADS superpotential as generated
by (fractional) instantons, although strictly speaking only the 3d ADS
superpotential has such a microscopic description.

It is interesting to point out that many of the manipulations in the
analysis of $\cn=1$ supersymmetric field theories usually carried out
in terms of the exact effective action, can be carried out
microscopically in terms of the physics of the relevant (possibly
fractional) instantons.  For instance, an important point in working
with gauge instantons in our examples below is the derivation, from
the instanton physics viewpoint, of the matching of scales in
processes like integrating out massive 4d matter etc. Let us describe
this in a simple example.  Consider an $SU(N_c)$ theory with $N_f<N_c$
flavours with mass (matrix) $m$, with dynamical scale
$\Lambda=(e^{-1/g^2})^{1/(3N_c-N_f)}$. Consider the situation where we
neglect the effect of the mass term on the instanton physics. Then the
instanton feels $N_f$ massless flavors and the non-perturbative
dynamics is described by the effect of a
$\frac{1}{N_c-N_f}$-fractional instanton, leading to the total
superpotential
\begin{eqnarray} W\, =\, (N_f-N_c)\, \left(\,
  \frac{\Lambda^{3N_c-N_f}}{\det Q{\tilde Q}}\,
\right)^{\frac{1}{N_c-N_f}}\, +\, m \, Q{\tilde Q}
\label{supomatch}
\end{eqnarray}

We may want to use an alternative description where we include the
effect of the mass terms from the start. From the spacetime viewpoint,
we integrate out the massive flavours. From the instanton perspective,
the instanton feels that the $2N_f$ fermion zero modes $\alpha,\beta$
associated to the flavors (in the D-brane picture, open strings
stretched between the instanton brane and the flavor branes) are
actually massive \footnote{This follows from the fact that the massive
  flavours are open strings from the color to the flavor branes, and
  that gauge brane instantons wrap exactly on top of the color
  branes.}. Integrating out the term $S_{\rm inst}=m \alpha \beta$ in
the instanton action leads to a prefactor of $\det m$ in the amplitude
for the left-over $\frac{1}{N_c}$-fractional instanton. Therefore the
superpotential is
\begin{eqnarray}
W\, = \, N_c\, \left(\, \Lambda^{3N_c-N_f}\,  \det m \,\right)^{\frac{1}{N_c}}
\label{supomatch2}
\end{eqnarray}
This is the standard $\frac{1}{N_c}$-fractional instanton amplitude for a SYM sector with effective scale $\Lambda'$ defined by
\begin{eqnarray}
\Lambda'^{\, 3N_c}\, =\,  \Lambda^{3N_c-N_f}\, \det m
\end{eqnarray}
Note that (\ref{supomatch2}) in fact agrees with (\ref{supomatch})
upon integrating out the massive flavours in the latter. Also, the
matching of scales is the familiar one in field theory.

\medskip

For future reference, let us mention that D-brane instantons
associated as above to 4d space filling D-branes, can lead to
non-perturbative superpotentials despite the fact that there are 4
universal zero modes in the instanton-instanton open string
sector. Indeed, two of these fermion zero modes have cubic couplings
to the bosonic and fermionic zero modes in the mixed open string
sector (strings stretched between the instanton and the gauge
D-branes). Their role can be regarded as imposing the fermionic
constraints to recover the ADHM instanton measure
\cite{Billo:2002hm}. The two left-over fermion zero modes are
Goldstinos of the $\cn=1$ supersymmetry, and are saturated by the
$d^2\theta$ integration involved in the induced superpotential.

\subsubsection{Non-gauge, ``exotic'' or ``stringy'' instantons}
\label{supostringy}

In general an euclidean D-brane instanton does not have the same
internal structure as any gauge D-brane in the configuration. Such
D-brane instantons do not have any known gauge field theory
interpretation, and are thus dubbed ``exotic'' or ``stringy''
instantons. BPS instantons of this kind lead to superpotential terms
only if they have two fermion zero modes, with additional fermion zero
modes forcing multi-fermion insertions leading to higher F-terms as
described below (additional fermions zero modes, with couplings to 4d
chiral multiplets, are regarded here as non-zero modes, since they are
lifted by background values of the latter; equivalently, integration
over these zero modes leads to insertions of the 4d chiral multiplet
in the induced superpotential). We are thus interested in stringy
instantons with two fermion zero modes.

In the same way as for gauge instantons, there are 4 universal fermion
zero modes in the instanton-instanton open string sector. However in
this case, there are no bosonic zero modes which can lift the two
non-goldstino modes. In the absence of other lifting mechanisms (like
closed string flux backgrounds), the only mechanism which can
eliminate these extra modes in type II perturbative models is an
orientifold projection. Therefore, only instantons invariant under the
orientifold action, and with a Chan-Paton action leading to an $O(1)$
symmetry, have two universal fermion zero modes, and have a chance of
leading to a non-perturbative superpotential (of course if they do not
have extra fermion zero modes in other sectors).

\subsubsection{Higher F-terms from D-brane instantons}

Besides D-brane instantons generating superpotentials, BPS D-brane
instantons with additional fermion zero modes lead to higher F-terms
in the effective action. These have been considered in
\cite{Beasley:2004ys,Beasley:2005iu}, and lead to operators with one
insertion of ${\ov D}{\ov \Phi}$ for each additional fermion zero
mode. Roughly speaking they have the structure
\begin{eqnarray}
\int \,d^4x\, d^2\theta\, w_{\ov i_1\ov j_1\ldots \ov i_n\ov j_n} (\Phi)\, 
{\ov D} {\ov \Phi}^{\ov i_1}\, {\ov D} {\ov \Phi}^{\ov j_1} \ldots
{\ov D} {\ov \Phi}^{\ov i_n}\, {\ov D} {\ov \Phi}^{\ov j_n}
\label{bwop}
\end{eqnarray}
where the tensor $w(\Phi)$ depends holomorphically on the 4d chiral
multiplets.  The simplest situation is an instanton with two
additional fermion zero modes, which is for instance realized for
gauge instantons in $N_f=N_c$ SQCD. The corresponding operator has the
above structure with $n=1$ and implements the familiar complex
deformation of the moduli space (in an intrinsic way, in the sense of
the moduli space geometry).

The study of the interplay between non-perturbative higher F-terms and
lines of marginal stability is beyond our scope in this paper,
although we expect that it admits a similar microscopic description in
terms of multi-instanton contributions after instanton splitting. In
any event, even for the analysis of superpotential terms, such
instantons will play an interesting role in some of our examples. We
refer to these instantons as Beasley-Witten instantons.

\subsection{A useful family of geometries}
\label{geometries}

Here we describe a set of geometries which we use in several of our
explicit examples below. They are non-compact geometries, but they
suffice to study instanton effects and transitions as long as they
involve just the local structure of compact cycles (see footnote
\ref{noncompact} for one example where non-compactness is relevant to
the discussion).

Let us consider the following class of local Calabi-Yau manifolds,
described by.
\begin{eqnarray}
xy= \prod_{k=1}^P (z-a_k) \nonumber \\
x'y'= \prod_{k'=1}^{P'} (z-b_k')
\end{eqnarray}
This kind of geometry is a particular case of those considered in
\cite{Ooguri:1997ih}. It describes two $\IC^*$ fibrations,
parametrized by $x,y$ and $x',y'$, varying over the complex plane $z$,
degenerating at the locations $a_i$, $b_i$ respectively. In this
geometry on can construct Lagrangian 3-cycles by considering segments
joining pairs of degeneration points on the base, and fibering the two
$\IS^1$'s in the two $\IC^*$ fibers. Segments joining pairs of
$a$-type degenerations or pairs of $b$-type degenerations lead to
3-cycles with topology $\IS^2\times \IS^1$. Segments joining $a$- and
$b$-type degenerations lead to 3-cycles with topology $\IS^3$. Let us
introduce the notation $[p_1,p_2]$ for the 3-cycle associated to the
pair of degeneration points $p_1$, $p_2$, whatever their type.

Introducing the holomorphic 3-form 
\begin{eqnarray}
\Omega \, = \, \frac{dx}{x} \, \frac{dx'}{x'}\, dz
\end{eqnarray}
the 3-cycle $[p_1,p_2]$ is calibrated by the form $e^{i\theta}\Omega$,
where $\theta$ is the angle of the segment $[p_1,p_2]$ with the real
axis in the $z$-plane. Namely $Im(e^{i\theta} \Omega)|_{[p_1,p_2]}=0$,
where $|_{[p_1,p_2]}$ denotes restriction to the 3-cycle. Hence,
segments which are parallel in the $z$-plane define 3-cycles which
preserve a common supersymmetry. We will be interested in
configurations where all degenerations are on (or near) the real axis.

We will consider stacks of 4d space filling D6-branes and/or euclidean
D2-branes wrapping the different 3-cycles, and describe the
non-perturbative superpotentials arising from these
configurations. The open string modes and their interactions are easy
to determine. For instance, each stack of $N$ D6-branes on a 3-cycle
leads to a $U(N)$ gauge group in a vector multiplet of ${\cal N}=1$
supersymmetry for 3-cycles of $\IS^3$ topology, and of ${\cal N}=2$
supersymmetry for 3-cycles of $\IS^2\times \IS^1$ topology. The angle
$\theta$ introduced above determines the precise supersymmetry
preserved by the corresponding set of branes. Also, two D6-branes
wrapping two 3-cycles involving one common degeneration point lead to
a vector-like pair of bi-fundamental chiral multiplets, arising from
open strings in the intersection of 3-cycles (which is topologically
$\IS^1$, coming from the $\IC^*$ that does not degenerate at the
intersection).

As discussed in \cite{Ooguri:1997ih} one can perform T-dualities along
the two $\IS^1$ directions, and map the configuration to a
Hanany-Witten setup of $P$ NS-branes (along 012345) and $P'$
NS'-branes (along 012389), with D4-branes (along 01236) suspended
among them, in a flat space geometry with a noncompact $x^6$ direction
(in contrast to the usual Hanany-Witten configurations describing
systems such as the conifold). The gauge theory content described
above follows from the standard rules in this setup (see
\cite{Giveon:1998sr}). This picture also facilitates the computation
of the superpotential, whose general discussion we skip, but which we
present in our concrete example below.

\section{Non-gauge D-brane instantons}
\label{sec:non-gauge}

In this section we consider ``exotic'' D-brane instantons
(i.e. instantons arising from D-branes wrapping internal cycles
different from those wrapped by the spacetime filling branes in the
model). For simplicity we restrict ourselves to perturbative type IIA
Calabi-Yau compactifications in the absence of fluxes. The aim of this
section is to show the continuity of the non-perturbative
superpotential across the lines of marginal stability for the
instantons. We show that the microscopic mechanism underlying this
continuity reveals interesting new properties of D-brane instanton
physics, including multi-instanton processes and non-perturbative
lifting of fermion zero modes.
 
We have already mentioned in Section \ref{supostringy} that in
perturbative type II models (and in the absence of additional
ingredients like 3-form fluxes), for instantons to have just the two
fermion zero modes required to contribute to the superpotential they
should be mapped to themselves under the orientifold action and have
an $O(1)$ Chan-Paton symmetry. This constrains the possible splittings
of the instanton in walls of marginal stability, for instance an
$O(1)$ instanton cannot split into two $O(1)$ instantons, as we show
in Appendix \ref{nosplit}. Still, there is enough freedom to have
non-trivial splitting of instantons that contribute to the
superpotential, as we now discuss in a simple example.

\subsection{$O(1)$ instanton splitting as $O(1)\times U(1)$ instantons}

\subsubsection{Configuration and marginal stability line}
\label{onetotwo}

In this section we consider one simple example of an $O(1)$ instanton
A, which contributes to the non-perturbative superpotential, and can
reach a line of marginal stability on which it splits as an O(1)
instanton $B$ and a $U(1)$ instanton (described as a brane $C$ and its
image $C'$).

\begin{figure}
\begin{center}
  
 \input{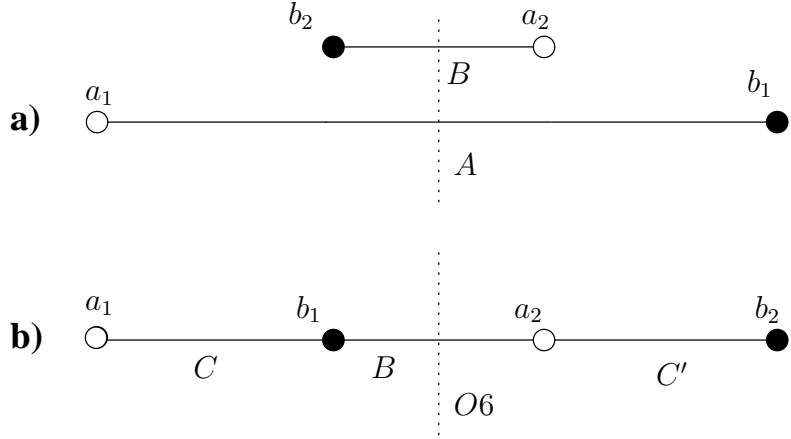}

  \caption{\small Example of an $O(1)$ instanton $A$ (figure a)
    splitting into an $O(1)$ instanton $B$ and a $U(1)$ instanton $C$
    and its image $C'$ (figure b).}
  \label{splitting1}
\end{center}
\end{figure}

Consider a geometry of the kind introduced in Section
\ref{geometries}, with two degenerations $a_1$, $b_1$ located at
$z=\pm t/2$, with $t\in \IR$, and two degenerations $a_2$, $b_2$
located at $z= \pm s/2+i\epsilon$, with $s, \epsilon \in \IR$, and
$s<t$ for concreteness, see Figure \ref{splitting1}. Namely
\begin{eqnarray}
x^2+y^2\, = \, (z+t/2)(z-s/2-i\epsilon) \nonumber\\
x'^2+y'^2\, =\, (z+s/2-i\epsilon)(z-t/2)
\end{eqnarray}
Consider modding out the geometry by the orientifold action $\Omega
R(-1)^{F_L}$, where $R$ is the antiholomorphic involution
\begin{eqnarray}
  z\to -{\ov z} \quad ; \quad (x,y) \leftrightarrow \left({\ov x}', {\ov y}'\right)
\label{orient}
\end{eqnarray}
The set of fixed points defines an O6-plane along the imaginary $z$
axis. This orientifold exchanges degenerations of $a$ and $b$
type. The parameters $s,t,\epsilon$ belong to chiral multiplets
associated to complex structure moduli invariant under the orientifold
action. We choose the O6-plane charge such that it leads to $O(1)$
Chan-Paton symmetry for D2-brane instantons on 3-cycles defined by
horizontal segments crossing it.

For generic non-zero $\epsilon$ there are two $O(1)$ instantons in
this configurations, corresponding to D2-branes on the segments
$[a_1,b_1]$ (denoted instanton A) and $[b_2,a_2]$ (denoted instanton
B). Each has just two fermion zero modes, and therefore leads to a
contribution to the non-perturbative superpotential
\begin{eqnarray}
W\, =\, f_1 e^{-T} \, +\, f_2 e^{-S}
\label{twoinst}
\end{eqnarray}
where $T,S$ are the closed string chiral multiplets whose real parts
are given by the moduli $t,s$ controlling the size of the wrapped
3-cycles. Here $f_i$ are prefactors given by one-loop determinants,
which depend on the Kahler moduli (but not on the complex structure
moduli).

When $\epsilon$ is taken to zero, the four degenerations align, and
the instanton A reaches a line of marginal stability, and splits into
an instanton of type $B$, and a $U(1)$ instanton corresponding to a
D2-brane on $[a_1,b_2]$ and its orientifold image on $[a_2,b_1]$
(denoted C and C' respectively). Since the complete superpotential
should behave continuously in this motion in moduli space, there
should be suitable instanton processes reproducing it. There are only
two basic instantons, namely the $O(1)$ instanton B on $[b_2,a_2]$,
which indeed reproduces the $e^{-S}$ term in (\ref{twoinst}), and the
$U(1)$ instanton C (with its image C'), which has four fermion zero
modes and does not contribute to the superpotential. Hence, there is
no instanton which reproduces the $e^{-T}$ term. In analogy with the
analysis in Section (\ref{gaugeinst}) for gauge D-brane instantons,
the resolution of the puzzle lies in understanding the mutual
influence of different instantons, and can be understood in different
ways as we now describe.

\subsubsection{The 2-instanton process}
\label{thetwoinst}

In order to show that the 2-instanton process contributes to the
superpotential, we have to discuss the structure of zero modes in the
2-instanton configuration, and how they are saturated. This will
involve the saturation of additional zero modes due to higher order
interactions on the instanton world-volume effective action.

Let us briefly describe the structure of zero modes in the different
sectors. We refer to the instantons $C$, $B$ as $1$, $2$ in this
section.

$\bullet$ In the 11 sector (and its $1'1'$ image), the open string
sector feels a background with 8 supercharges, half of which are
broken by the instanton. We have a $U(1)$ gauge symmetry (although
there are no gauge bosons), four bosonic zero modes $x_1^\mu$
corresponding to the 4d translational Goldstones, and four fermionic
zero modes $\theta_1^\alpha$, ${\tilde \theta}_{1\,\dot{\alpha}}$,
corresponding to the Goldstinos. Note that the Lorentz symmetry under
which these are chiral spinors is a global symmetry from the instanton
volume viewpoint.

$\bullet$ The 22 sector is sensitive to the orientifold action and
hence feels a background with 4 supercharges, half of which are broken
by the instanton. The orientifold projection truncates part of the
spectrum, as compared with the above $U(1)$ instanton case. There is
an $O(1)\equiv \IZ_2$ gauge symmetry, and four bosonic zero modes
$x_2^\mu$.

$\bullet$ Consider now the spectrum from open string stretching at the
12 intersection (and its image 1'2). Locally around it, the background
admits 16 supersymmetries, half of which are broken by the
D-branes. The massless modes thus form a hypermultiplet under the
unbroken 8 supersymmetries. We have two complex bosonic zero modes
$\varphi_{12}$, $\varphi_{21}$, with charges $+1$ and $-1$ under the
$U(1)_1$ gauge symmetry of the instanton 1, and four fermionic zero
modes, $\chi_{12}^\alpha$, $\chi_{21}^\alpha$, with charges $+1$ and
$-1$ under $U(1)_1$. Alternatively, these can be conjugated to ${\ov
  \chi}_{21\, \dot{\alpha}}$, ${\ov \chi}_{12\, \dot{\alpha}}$, with
charges $-1$, $+1$. Let us call the chiral superfields in the
hypermultiplet $\Phi_{12}$ and $\Phi_{21}$.

\medskip

Let us now describe the couplings of these modes on the volume of the
instanton. They are analogous (upon dimensional reduction) to the couplings that would appear if we would have D6-branes instead of D2-branes. There is a first term which describes the mass terms of the open strings between the two instantons when
they are separated in the 4d direction
\begin{equation}
  S_{kinetic}\, =\, (x_1^\mu - x_2^\mu)^2\, (|\varphi_{12}|^2 +
  |\varphi_{21}|^2)\, +\,i(x_1^\mu - x_2^\mu)\, \{\, {\ov \chi}_{12} \sigma_\mu \chi_{12} -
  {\ov \chi}_{21} \sigma_\mu \chi_{21}\,\}
  \label{eq:kinetic-couplings}
\end{equation}
These terms are related to the couplings to gauge bosons in the D6-D6
system. There are also terms involving the neutral fermion zero modes
$\theta$, ${\tilde \theta}$ (analogous to the couplings to gauginos in
the D6-D6 system), given by \footnote{Similar couplings in the context
  of a D2-instanton intersecting its orientifold image have been
  described in \cite{Blumenhagen:2007bn}.}.
\begin{equation}
  S_{\lambda} = (\chi_{12}\,(\theta_1 - \theta_2))\varphi_{12}^* -
  (\chi_{21}\,(\theta_1 - \theta_2))\varphi_{21}^* + ({\ov
    \chi}_{12}\tilde\theta)\varphi_{12} - ({\ov
    \chi}_{21}\tilde\theta)\varphi_{21}
  \label{eq:gaugino-couplings}
\end{equation}
Notice that the combination $\theta_1+\theta_2$ is decoupled, and
corresponds to the two Goldstinos of the combined two-instanton
system. We also have a D-term potential (the same arising in a
D6-D6-brane system):
\begin{equation}
S_D\, =\, (\, |\varphi_{12}|^2 - |\varphi_{21}|^2\, )^2
\end{equation}
Finally, there are quartic couplings involving the fields in the 12
sector. The local intersection preserves 8 supercharges, but the
interaction is induced by effects that preserve only 4 supercharges
(due to the different nature of degenerations at the intersection and
adjacent to it). The interaction can be obtained from a superpotential
of the form
\begin{eqnarray}
W\, \simeq \, (\, \Phi_{12}\Phi_{21}\,)^2
\end{eqnarray}
This in fact identical to the superpotential that would be obtained
for D6-branes. The underlying reason is that both D2- and D6-branes
have identical boundary states of the internal CFT (and a flip of DD
to NN boundary conditions in the 4d part), thereby leading to
essentially the same correlation functions.

Thus we obtain fermion-scalar interactions of the form
\begin{eqnarray}
  S_{\chi^2\varphi^2}\, =\, \chi_{12}\varphi_{21}\chi_{12}\varphi_{21} \, +\, 
  2\chi_{12}\chi_{21}\varphi_{12}\varphi_{21} \, +\,
  \varphi_{12}\chi_{21}\varphi_{12}\chi_{21} \quad + \text{h.c.}
\end{eqnarray}
and the F-term scalar potential
\begin{eqnarray}
  S_{F}\, =\, |\varphi_{21}\varphi_{12}\varphi_{21}|^2\, +\, |\varphi_{12} \varphi_{21}\varphi_{21}|^2
\end{eqnarray}

Let us now consider the role of this complete instanton effective
action in the generation of a non-perturbative superpotential. Notice
that the contribution to the superpotential is dominated by
configurations of overlapping instantons, namely when $x_1-x_2=0$, as
follows. A large non-zero separation gives large masses to the open
strings between the instantons (consistent with the equation
(\ref{eq:kinetic-couplings})), so we can integrate out these fields
and set their vevs to zero, making the couplings in
(\ref{eq:gaugino-couplings}) vanish. Then we cannot saturate the
$\theta_1-\theta_2$ zero modes, and the integral vanishes. So let us
focus for simplicity \footnote{In fact it is possible, and not much
  harder, to carry out the computation allowing for arbitrary
  $x_1-x_2$. Namely one can perform the Gaussian integration over
  these bosonic zero modes, and conclude that the result is localized
  (with some exponentially vanishing tail) onto $x_1=x_2$. We omit the
  detailed analysis since the conclusions are essentially unchanged,
  and the simplified discussion is enough to show that the 2-instanton
  process at hand provides a non-trivial contribution.} in the case
$x_1-x_2=0$. In this case we have an instanton action given by
$S_{\text{2inst}}=S_\lambda+S_D+S_{\chi^2\varphi^2}+S_F$. The pieces
relevant for the saturation of zero modes will be $S_\lambda$ and
$S_{\chi^2\varphi^2}$. We can soak up $(\theta_1-\theta_2)$ by
bringing down two insertions of $(\chi_{12}\,(\theta_1
-\theta_2))\varphi_{12}^*$ from $S_\lambda$. Similarly we can soak up
$\tilde\theta$ by bringing down two insertions of $({\ov
  \chi}_{12}\tilde\theta)\varphi_{12}$. This also saturates the zero
modes $\chi_{12}$, ${\ov \chi}_{12}$. The remaining zero modes
$\chi_{21}$, ${\ov \chi}_{21}$ can be soaked up by bringing down two
insertions of $\varphi_{12}\chi_{21}\varphi_{12}\chi_{21}$ from
$S_{\chi^2\phi^2}$ and two insertions of its complex conjugate
operator. Bringing everything together, and integrating over the
(saturated) fermionic zero modes, we get the following 2-instanton
contribution:
\begin{equation}
  \int d^4x_+ d^2\theta_+ [d\varphi]\, 
  \exp\left\{-S_D-S_F\right\} |\varphi_{12}|^4
\end{equation}
where $x_+=x_1+x_2$, $\theta_+=\theta_1+\theta_2$ are the surviving
zero modes of the instanton. Note that the $\varphi$ integral
converges since there are no flat directions in the
($\varphi_{12},\varphi_{21}$) space, as is easily seen from the form
of $S_D$ and $S_F$.  There are other similar contributions from other
combinatorics of soaking up zero modes. The overall result is a
non-zero contribution to the superpotential from the 2-instanton
process.

The above mechanism is very similar to the lifting of accidental zero
modes by world-volume interactions in other situations. For instance
in the study of instanton effects on 4d $\cn=4$ supersymmetric
theories, where a world-volume 4-fermion interaction lifts fermion
zero modes in groups of four (and allows multi-instanton processes
contribute to the same 4d effective action terms as single-instanton
ones). The analogy could be made much more explicit by integrating
over the bosonic modes above, generating world-volume 4-fermion
interactions. This is, to our knowledge, the first explicit
realization of a similar mechanism in the computation of
non-perturbative D-brane instanton superpotentials in $\cn=1$
theories. Notice also the interesting fact that in such situations the
usual recipe of adding the contributions from the individual
instantons misses these new contributions.

\begin{figure}
\begin{center}
  
 \input{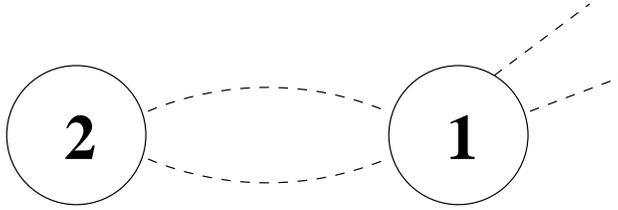}

  \caption{\small Schematic picture of a multi-instanton configuration
    contributing to the superpotential. A number of additional fermion
    zero modes are saturated against each other, due to interaction
    terms in the world-volume effective action of the 2-instanton
    system. The two left-over fermion zero modes are the Goldstinos of
    the overall BPS D-brane instanton system, and are saturated
    against the $d^2\theta$ integration in the induced 4d effective
    action superpotential term.}
  \label{twoinstanton}
\end{center}
\end{figure}

The spacetime picture of the above mechanism is of the kind shown in
Figure \ref{twoinstanton}, with two fermion zero modes of each
instanton saturated against each other, and two left-over fermion zero
modes.

As a last comment, note that the above system fits nicely with the
concept of quasi-instanton as described in \cite{Dorey:2002ik}. Namely
the bosonic modes $\varphi$ can be described as quasi-zero modes, and
they parametrize a quasi-moduli space of quasi-instantons, in the
sense that they correspond to a moduli space of instantons, which are
lifted by a world-volume potential whose effects can be studied
perturbatively in the value for the bosonic fields. Although strictly
speaking such configurations do not correspond to BPS instantons, they
can provide the dominant dynamical effect in the semiclassical
approximation to certain quantities. Note that the additional
Goldstinos (those associated to the supersymmetries preserved by BPS
instantons) are not turned on in the first correction, and the effect
of larger values for the bosonic fields is suppressed due to the
exponential damping.

\subsubsection{Non-perturbative lifting of zero modes of the $U(1)$ instanton}
\label{nplift}

One can interpret the appearance of the non-trivial contribution to
the superpotential as the instanton 2 generating an effective
interaction term for the additional zero modes of the instanton
1. Indeed the piece
\begin{eqnarray}
  \Delta S_{\rm inst 1} = \int \, d^2\theta_2 \, d^4\chi \, d^4\varphi \,
  \exp [\, (\theta_1-\theta_2) \, \varphi \, \chi\,
  + \, {\tilde \theta}_1\,  {\ov \varphi}\,  {\ov \chi} +  \chi^2\, \varphi^2\,  +\,  V(\varphi)\,]
\end{eqnarray}
of the integral above can be regarded as computing the
non-perturbative contribution of the instanton 2 to the effective
action of the instanton 1. The result corresponds to an effective mass
term (of non-perturbative strength $e^{-S}$) for the extra fermion
zero modes of the instanton 1. Hence the amplitude of the instanton 1
is sketchily of the form
\begin{eqnarray}
  S_{4d} & \simeq & \int d^4x\, d^2\theta \, d^2{\tilde \theta}\, \exp
  \, (\,-T_1\, -\, e^{-S} \, {\tilde \theta} {\tilde \theta} )
  \notag\\
  & = & \int d^4 x\, d^2\theta\, e^{-S} e^{-T_1}\, = \,\int\, d^4x\,
  d^2\theta e^{-T}
  \label{twoexpo}
\end{eqnarray}
namely the appropriate superpotential term. 

In Section (\ref{npliftgauge}) we will provide yet another viewpoint
regarding the non-perturbative lifting of fermion zero modes.

It is very interesting that $U(1)$ instantons can contribute to
non-perturbative superpotentials via this mechanism of
non-perturbative lifting of the extra zero modes. We also expect other
instantons with additional universal fermion zero modes, like $Sp(2)$
instantons, to similarly contribute under special circumstances. It
would be interesting to use this mechanism to revisit the role of
interesting $U(1)$ and $Sp(2)$ instantons in model building
applications, like the instanton scan in \cite{Ibanez:2007rs}. In
fact, multi-instanton processes can already arise in simple toroidal
orientifolds (see \cite{Ibanez:2007tu} for an explicit $\IT^6/\IZ_3$
example).

\subsubsection{4d charged matter insertions}
\label{insertions}

The bottom line of the above Sections is that non-perturbative
superpotentials for non-gauge D-brane instantons are continuous across
lines of marginal stability. The microscopic instanton physics
mechanism relies on the fact that additional zero modes in
multi-instanton processes can be saturated by interactions, leaving
only a few zero modes to be saturated by external insertions in 4d
correlators. The initial instanton amplitude is thus fully
reconstructed by a multi-instanton amplitude.

Let us comment on the situation where the initial instanton intersects
some of the 4d space filling D-branes in the system. There are fermion
zero modes charged under the 4d gauge group at those intersections. In
order to contribute to the superpotential, these additional fermion
zero modes should be coupled to operators involving the 4d charged
matter fields, so that upon integration over them (or pulling down
these interactions) one generates insertions of the 4d charged matter
fields in the 4d effective superpotential as discussed in
\cite{Ganor:1996pe,Blumenhagen:2006xt,Ibanez:2006da,Florea:2006si}. The
appearance of the same insertions in the multi-instanton amplitude at
the line of marginal stability is easy to show: Notice that the
homology charge of the contributing D-brane instanton system is
preserved in the process of reaching the line of marginal
stability. This ensures that the number of charged fermion zero modes
is preserved in the process, and that the insertions of 4d fields are
suitably generated. We refrain from delving into a more detailed
discussion in concrete example, and prefer to move on.

\subsection{$O(1)$ splitting as $U(1)$ instanton}
\label{onetoone}

In this Section we would like to consider another possible splitting
of an $O(1)$ instanton across a line of marginal stability, in which
it splits as a $U(1)$ instanton and its image. In fact this kind of
process was considered in \cite{Blumenhagen:2007bn}, with the
conclusion that such instantons cannot contribute to the
superpotential due to the presence of additional zero modes. In fact
our explicit example evades this no-go result: there exists an F-term
interaction in the world-volume of the instanton (not considered in
\cite{Blumenhagen:2007bn}) which lifts the additional fermion zero
modes.

\begin{figure}
\begin{center}
  
 \input{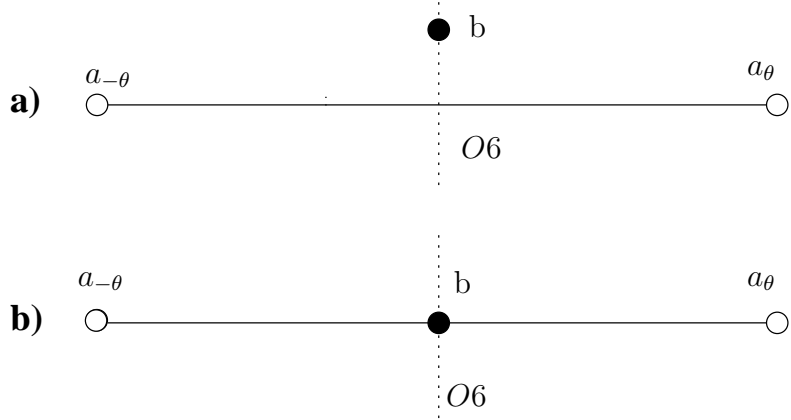}

  \caption{\small Configuration of an $O(1)$ instanton splitting as a
    $U(1)$ instanton (and its orientifold image). Interpreted as a HW
    setup, the dots $b$, $a_{\pm \theta}$, denote the locations in the
    67 plane for an unrotated NS-brane, and NS5-branes rotated by
    angles $\pm \theta$ in the 4589 directions. Interpreted as
    D1-brane instantons in a threefold geometry, the dots $a_{\pm
      \theta}$, $b$ denote a projection of the degenerations loci of a
    $\IC^*$ fiber. D1-brane instantons wrap 2-cycles obtained by
    fibering the latter over segments defined by such degenerations,
    and are supersymmetric when the segments lie horizontally.}
  \label{unfoldgeom}
\end{center}
\end{figure}

The geometry in this configuration is similar, but slightly different
from those introduced in Section \ref{geometries}. It is therefore
better to introduce the configuration in terms of a type IIA
Hanany-Witten setup. Consider a NS-brane along 012345, and two
NS-branes along 0123 and rotated by angles $\theta$ and $-\theta$ in
the planes 45 and 89 (so we denote them NS$_\theta$ and
NS$_{-\theta}$). One can discuss the relevant part of the geometry by
depicting the positions of the different branes in the $z=x^6+ix^7$
plane, as shown in Figure \ref{unfoldgeom}. In our configuration,
the NS-brane is located at $z=i\epsilon$, while the NS$_{\pm
  \theta}$-branes are located at $z=\pm t$, with
$t,\epsilon\in\IR$. We consider instantons arising from euclidean
D0-branes suspended between the different NS5-branes, thus
corresponding to segments between the different NS5-brane locations in
the $z$-plane. BPS instantons correspond to horizontal segments.

The above kind of configuration can be T-dualized using
\cite{Uranga:1998vf} into a type IIB geometry similar to those in
Section \ref{geometries} (and similar to those studied in
\cite{Gubser:1998ia,Lopez:1998zf}). As a complex variety, the geometry
can be described as an unfolding of an $A_2$ singularity
\begin{eqnarray}
xy= u (u+\alpha v)(u-\alpha v)
\label{unfolding}
\end{eqnarray}
with $\alpha=\tan\theta$. It can be regarded as a $\IC^*$ fibration
over the $(u,v)$ space, and degenerating at the loci $u=0$, $u=\pm
\alpha v$. The directions $u$, $v$ are closely related to the
directions $45$ and $89$ in the HW setup, and the degeneration loci
correspond to the NS5-brane volumes in those directions.  The geometry
contains non-trivial 2-cycles, obtained by fibering the circle in the
$\IC^*$ over a segment joining two degeneration loci. There are
D-brane instantons arising from D1-branes wrapping these 2-cycles.
The description of the geometry as a complex manifold provided in
(\ref{unfolding}) does not encode the parameters $\epsilon, t$, which
are Kahler parameters and control the lines of marginal stability of
our instantons. We will rather use pictures like Figure
\ref{unfoldgeom}, which can be regarded as a depiction of the blow-up
structure of the above geometry, or the representation of the $67$
plane in the HW configuration.  Since the spectrum of instanton zero
modes and their interactions can be obtained from the latter using
standard rules, we stick to this language, although it is
straightforward to translate into the geometric one.

Let us introduce an O6-plane along 0123789 in the HW setup, which thus
corresponds to a fixed line along the vertical axis on the
$z$-plane. The O6-plane intersects the NS-brane (in an intersection
preserving 8 supercharges) mapping it to itself, while it exchanges
the NS$_{\pm \theta}$-branes. We choose the O6-plane charge such that
it leads to $O(1)$ Chan-Paton symmetry on instantons along horizontal
segments crossing the O6-plane.

Consider the configuration for non-zero $\epsilon$, see Figure
\ref{unfoldgeom}a. The only BPS instanton is given by a D0-brane
stretched between the NS$_{-\theta}$ and NS$_\theta$ branes. It has
$O(1)$ Chan-Paton symmetry and has just 2 fermion zero modes (for
non-zero $\theta$), and thus leads to a non-perturbative
superpotential contribution $W \simeq e^{-T}$, with $T$ the chiral
multiplet with real part $t$.

Consider the configuration for $\epsilon=0$, where the previous
instanton reaches a line of marginal stability and splits into a
$U(1)$ instanton $1$ (a D0-brane between the NS$_{-\theta}$ and the
NS branes) and its orientifold image $1'$ (between the NS and
NS$_\theta$ branes). At the Gaussian level, the instanton has many
additional zero modes beyond the required set of two fermion zero
modes, hence naively it would not contribute to the
superpotential. However, it is easy to go through the analysis of zero
modes and their interactions, and realize that the additional fermion
zero modes are lifted. The argument is very similar to that in
previous Section, so our discussion is sketchy.

In the $11$ sector of open strings with both endpoints on the
instanton, there are four translational Goldstone bosonic zero modes
$x^\mu$, and four fermionic zero modes, two of them $\theta^\alpha$
associated to Goldstinos of the 4d $\cn=1$, and two
${\tilde\theta}_{\dot \alpha}$ associated to the accidental
enhancement to $\cn=2$. In the $11'$ sector of open strings between the
instanton and its image, we have a hypermultiplet (given by the pair
of chiral field $\Phi$ and $\Phi'$ in $\cn=1$ language) of zero modes
$\varphi$, $\varphi'$, $\chi_\alpha$, $\chi'_\alpha$ with $U(1)$
charges $\pm 2$ for unprimed/primed fields. The couplings between the
$11$ and $11'$ fields are
\begin{eqnarray}
  S\, =\, {\tilde \theta} \, (\, \varphi {\ov \chi}  - {\ov \chi}' \varphi'\, )
\end{eqnarray}
From the HW construction it is possible to derive that there are
interactions among fields in the $11'$ sector. Given the amount of
susy, it is possible to describe them by a superpotential $W\simeq
(\Phi \Phi')^2$. Namely, there are scalar potential terms (involving
also a D-term contribution)
\begin{eqnarray}
  V_D & \simeq  &  (\, |\varphi|^2\, -\, |\varphi'|^2\, )^2  \nonumber \\
  V_F & \simeq & |\varphi\varphi'\varphi|^2 \, + \, |\varphi'\varphi\varphi'|^2
\end{eqnarray}
and most importantly couplings to the $11'$ fermions
\begin{eqnarray}
  S_{\varphi\chi} \, \simeq\, \chi\chi\varphi'\varphi' \, +\, 2\, \chi \varphi\chi'\varphi' \, +\, \varphi\varphi\chi'\chi'
\end{eqnarray}
As discussed in previous examples, all additional zero modes can be
saturated by pulling down interaction terms from the instanton
effective action. The only left-over fermion zero modes are the two
Goldstinos $\theta^\alpha$, hence the $U(1)$ instanton contributes to
the superpotential. Note that in contrast with the previous examples,
the lifting of zero modes of the $U(1)$ instanton is purely
perturbative (although is reminiscent of the non-perturbative lifting
in previous section when regarded in the covering space).

Since the volume of the instanton and its image add up to the volume
of the original $O(1)$ instanton, the complete superpotential is
continuous.

\section{Gauge D-brane instantons}
\label{gaugeinst}

Let us proceed to systems which are more familiar, namely
configurations where the non-perturbative superpotential can be
regarded as generated by gauge theory instantons. The idea is to
consider a simple example of gauge sector with a non-perturbative
superpotential, engineered via D-branes, and to consider its fate as
one crosses a line of marginal stability. The general lesson of this
example is the following.  In this kind of setup, the crossing of
lines of marginal stability in moduli space is basically described in
terms of a Higgsing/unHiggsing in the field theory. Also, the
dependence of the superpotential on the relevant moduli is encoded in
the dynamical scales of the gauge factors associated to the 4d
spacetime filling D-branes (since they control the gauge
couplings). Thus the statement about the continuity of the
superpotential across lines of marginal stability corresponds to the
familiar matching of dynamical scales of a gauge theory in a
Higgsing/unHiggsing process, at energies above and below the relevant
vevs. Given this interpretation and the construction and discussion
below, it is easy to find other examples of similar behaviour.

\subsection{An example of $N_f<N_c$ SQCD non-perturbative superpotential}
\label{sqcd}

\subsubsection{Configuration, marginal stability, and the spacetime picture}

\begin{figure}
\begin{center}
  
 \input{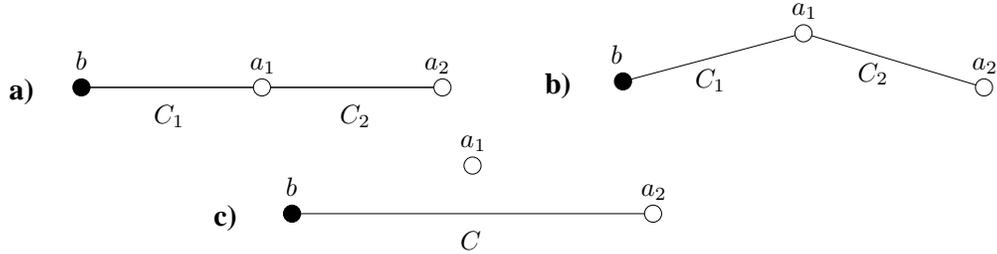}

  \caption{\small a) Marginally stable configuration. b) Moving $a_1$
    away from the horizontal axis renders the configuration
    nonsupersymmetric, so it can c) decay to a supersymmetric
    configuration by brane recombination.}
  \label{crossgauge}
\end{center}
\end{figure}

Let us describe a system of D6-branes crossing a line a marginal
stability in a geometry of the kind introduced in Section
\ref{geometries}. Consider the geometry in Figure~\ref{crossgauge},
having two $a$-type degenerations and one $b$-type degeneration,
ordered as $b,a_1,a_2$ from left to right along the real axis. We
consider a set of $N$ D6-branes wrapped on the 3-cycle $C_1=[b,a_1]$
and $N$ D6-branes on $C_2=[a_1,a_2]$. This configuration is
supersymmetric as long as the degeneration $a_1$ is aligned with the
other two. Moving $a_1$ away from the horizontal axis forces the
D6-branes on $C_1$ and $C_2$ to misalign, and their tension
increases. The system of branes can relax by forming a bound states,
described by $N$ D6-branes on the 3-cycle $C=[b,a_2]$. Namely, the
locus in moduli space where $a_1$ aligns with $b,a_2$ corresponds to a
line of marginal stability for a D6-branes on $C$, which become
unstable against decay into D6-branes on $C_1$, $C_2$.

The above phenomenon of brane dynamics has a counterpart in classical
gauge field theory. The system of $N$ D6-branes on $C_1$, $C_2$ leads
to a $U(N)_1\times U(N)_2$ ${\cal N}=1$ supersymmetric gauge theory
(see later for a discussion of the $U(1)$ factors), with chiral
multiplets $Q$, ${\tilde Q}$ in the $(\fund_1,\antifund_2)$,
$(\antifund_1,\fund_2)$, and $\Phi$ in the adjoint of $SU(N)_2$. There
is a classical superpotential
\begin{eqnarray}
W\, =\, \tr \, Q\Phi{\tilde Q}
\end{eqnarray}
The parameters of the gauge theory are the gauge couplings $g_i$, and
theta angles $\theta_i$, which are classically related to $C_i$ by
\begin{eqnarray}
T_i\, =\, \frac{1}{g_i^{\, 2}} +i\theta_i\, =\,\frac{1}{g_s} \int_{C_i}\, \Omega \, +\, i \int_{C_i}\, A_3
\end{eqnarray}
where $A_3$ is the type IIA RR 3-form. In the quantum theory, these parameters are traded for dynamical scales 
\begin{eqnarray}
\Lambda_1\, =\, \exp\left(-\frac{T_1}{2N}\right) \quad ; \quad 
\Lambda_2\, =\, \exp\left(-\frac{T_2}{N}\right)
\end{eqnarray}

The change in the complex structure associated to moving $a_1$ off the
horizontal axis corresponds to turning on a Fayet-Iliopoulos parameter
$\xi$ for the difference of the two $U(1)$'s. This triggers a vev for
the bi-fundamental flavours $Q$ or ${\tilde Q}$, depending on the sign
of $\xi$, and breaking the gauge group to the diagonal
$U(N)$. Assuming that $Q$ acquires the vev, the fields $\Phi$,
${\tilde Q}$ become massive by the superpotential and disappear. We
are left with a $U(N)$ pure SYM gauge theory, with complex gauge
coupling
\begin{eqnarray}
T\, =\, T_1+T_2\, =\, \int_C\, \Omega \, +\, \int_C\, A_3 
\end{eqnarray}
This agrees with the picture of the D6-branes recombining into
D6-branes wrapped on $C$.

It is worth noting that the $U(1)$ generators have $BF$ Stuckelberg
couplings with closed string moduli, which make the gauge bosons
massive, so the $U(1)$ factors are really absent from the low energy
effective theory. This modifies the above discussion very
mildly. Namely, instead of turning on a FI parameter, the above
transition can be regarded as moving along the baryonic branch of the
$SU(N)_1\times SU(N)_2$ theory, to yield a pure $SU(N)$ SYM theory.

\medskip

We would now like to consider the non-perturbative superpotential in
these two systems, and in showing that it is continuous across the
line of marginal stability. Interestingly, the non-perturbative
effects have a microscopic description in terms of D2-brane instantons
on the relevant 3-cycles, along the lines described in Section
\ref{introinst}. In the discussion we stick to the description in
gauge theory language. Also for convenience we use the description
where the $U(1)$'s are not included in the low-energy dynamics.

Consider first the system of $N$ D6-branes on $C$. Since it
corresponds to a pure SYM theory, it confines and develops a gaugino
condensate. There is a non-perturbative superpotential
\begin{eqnarray}
W\, =\, \Lambda^3\, =\, (\, e^{-T/3N}\,)^3
\label{supo1}
\end{eqnarray}

Consider now the situation when the instanton reaches the line of
marginal stability. We consider the system of $N$ D6-branes on $C_1$
and $N$ D6-branes on $C_2$, so we essentially have to study the
dynamics of the $SU(N)_1\times SU(N)_2$ gauge theory. Let us focus in
the regime where $\Lambda_1\gg \Lambda_2$, so the dynamics of
$SU(N)_1$ dominates.

In this case the $SU(N)_1$ group confines first. It has $N_f = N_c$,
so the instanton on $C_1$ is a Beasley-Witten instanton, which induces
a quantum deformation on the moduli space. Instead of using the
intrinsic picture in moduli space and inducing an operator of the form
(\ref{bwop}), we prefer to work as usual in field theory analysis, by
imposing the deformation by a quantum modified constraint. We describe
the system is in terms of mesons $M$ and baryons $B\tilde B$, with
superpotential:
\[
W = \mu \Phi M + \mu^{-2N+2}\, X\, ( \det M - B \tilde B - \Lambda_1^{2N})
\]
where we have introduced the scale $\mu$ to keep the dimension of the
operator in the superpotential invariant. This dynamical scale will be
of the order of $\Lambda_1$, so we use it in what follows.

The F-term for $\Phi$ enforces $M=0$, and vice versa. The fields
$\Phi$ and $M$ are massive, so we can integrate them out. We are left
with a pure $SU(N)_2$ SYM theory, with dynamical scale $\Lambda_f$, to
be determined later. In addition we have the singlets $X$, $B$,
${\tilde B}$, with superpotential
\[
W \simeq X\,(\,B\tilde B + \Lambda_1^{2N}\, )
\]
The theory has a one-complex dimensional baryonic moduli space, but
these singlets do not modify the theory otherwise.

The dynamical scale $\Lambda_f$ is determined by the matching, in
analogy with the discussion in (\ref{supogauge}), as
\begin{eqnarray}
\Lambda_f^{3N}=\Lambda_2^{N}\Lambda_1^{2N}
\end{eqnarray}
and is in fact the same as the $\Lambda$ introduced above. 

In this left-over $SU(N)_2$ pure SYM theory, the effect of the
(fractional) instanton on $C_2$ is simply to develop a gaugino
condensate non-perturbative superpotential
\begin{eqnarray}
W\, =\, \Lambda_f^{\, 3} \, =\, (\, e^{-T/3N}\, )^3
\end{eqnarray}
in agreement with (\ref{supo1}).

This example provides a non-trivial and simple realization of the
continuity of superpotentials across lines of marginal stability.  The
instanton wrapping $C$ reaches the line of marginal stability, at
which it splits into two BPS instantons, wrapping $C_1$ and $C_2$.
The instanton on $C_1$ is of Beasley-Witten type and deforms the
moduli space. The instanton on $C_2$, once the effect of the instanton
on $C_1$ is taken into account, induces a non-perturbative
superpotential. The total effect neatly adds up to the effect of the
single instanton on $C$ before crossing the line of marginal
stability.

For completeness, let us mention that the discussion with $U(1)$'s in
the effective action is similar. There are no baryonic operators, so
there are no fields left out after integrating out $M$, $\Phi$. The
one-dimensional moduli space is realized in this view in the closed
sector, as the FI term for the relative $U(1)$ corresponding to the
position of $a_1$ off the horizontal axis.

\subsection{Microscopic interpretation}
\label{microgauge}

In this Section we discuss the microscopic interpretation of the
continuity of the non-perturbative superpotential of the above
configuration in terms of D-brane instanton physics.

\subsubsection{The 2-instanton process}

In analogy with the discussion for non-gauge instanton in Section
\ref{thetwoinst}, and from the above discussion, it is clear that the
superpotential contribution at the line of marginal stability arises
from a two-instanton process, involving the instantons $C_1$ and
$C_2$. In fact, it is possible to compute the set of zero modes for
the two-instanton system, and their interactions.

We skip the detailed discussion and just sketch the result. The
contributions to the superpotential localize on configurations of
instantons coincident in 4d. In addition the 3-cycle $C_1$ is
non-rigid, and there is a bosonic zero mode $\phi$ parametrizing a
branch where the instanton on $C_1$ slides away from the D6-branes on
$C_1$. Along this branch the configuration has additional zero modes
$\chi$, ${\ov \chi}$ (the partners of $\phi$), which are not
saturated. Hence the contributions to the superpotential localize at
$\phi=0$. At this point one can easily check that all fermion zero
modes except for the two overall Goldstinos $\theta_1+\theta_2$ have
non-trivial interactions, which can be pulled down to saturate the
corresponding integrals.

\begin{figure}
\begin{center}
  
 \input{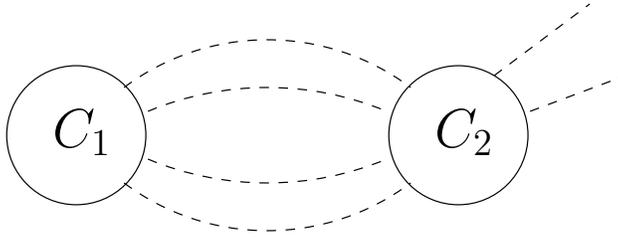}

  \caption{\small Schematic picture of the multi-instanton
    configuration discussed in the text.}
  \label{twoinstantongauge}
\end{center}
\end{figure}

The whole process can be described in spacetime in terms of a diagram
\ref{twoinstantongauge}. The instanton $C_2$ has six unsaturated
fermion zero modes, since it is a Beasley-Witten instanton with
$N_f=N_c$ (thus leading to two unsaturated fermion zero modes beyond
the two $\cn=1$ Goldstinos) and two additional fermion zero modes
$\chi$, ${\ov chi}$ from being on a non-rigid cycle. The instanton
$C_1$ has four unsaturated fermion zero modes, since it is an
$N_f=N_c$ Beasley-Witten instanton. In the two-instanton process, one
can generate interactions between the zero modes of the two instantons
via the bosonic zero modes charged under both, which allow to contract
four fermion zero modes, leading to an overall process with only two
fermion zero modes.

\subsubsection{Non-perturbative lifting of fermion zero modes}
\label{npliftgauge}

Along the lines of the discussion in Section \ref{nplift}, we would
like to improve on the additional viewpoint of the process as a
lifting of fermion zero modes of the instanton $C_2$ by a
non-perturbative effect induced by the instanton $C_1$. In fact, for
gauge instantons the mechanism can posed in a much sharper
setup. Consider a gauge instanton $A$, with field configuration ${\cal
  A}(x^\mu;\varphi,\psi)$, as a function of the sets of bosonic and
fermionic zero modes, $\varphi$, $\psi$. Here ${\cal A}$ denotes the
set of all 4d fields involved in the configuration. The classical
effective action for the zero modes $S_{\rm inst.}(\varphi,\psi)$ can
be obtained by replacing the instanton field configuration on the 4d
action $S_{4d}[{\cal A}]=\int d^4 x \, {\cal L} [{\cal A}]$, namely
\begin{eqnarray}
  S_{\rm inst}(\varphi,\psi)\, =\, \int d^4x\, {\cal L} [{\cal A} (x^\mu; \varphi,\psi)]
\end{eqnarray}
From this point of view, any additional term in the 4d effective
action $\delta S_{4d}[\cal A]$ induces a corresponding term on the
instanton effective action $\delta S_{\rm inst}(\varphi,\psi)$.
\begin{eqnarray}
  \delta S_{\rm inst} \, =\, \int d^4 x\, \delta {\cal L}[{\cal A}
  (x^\mu;\varphi,\psi)]
\end{eqnarray}
When the additional term in the 4d effective action $\delta S_{4d}$ is
induced by another instanton $B$, the term $\delta S_{\rm
  inst.}(\varphi,\psi)$ can in a very precise sense be regarded as a
non-perturbative interaction term for the zero modes of $A$ induced by
the instanton $B$. In particular interaction terms of this kind
involving the fermion zero modes $\psi$ of $A$ are non-perturbatively
lifted by the instanton $B$. Notice that the $g_s$ dependence arranges
in such a way that the total 4d effect is suppressed by the
exponential factors of both instantons $A$ and $B$, as in
(\ref{twoexpo}).

On general grounds, we may expect that an instanton $B$ with $k$
fermion zero modes induces a 4d F-term leading to contributions to
$S_{4d}$ with $k$ 4d fermion insertions. This will in general induce
an interaction term $S_{\rm inst.}(\varphi,\psi)$ on the instanton $A$
lifting $k$ fermion zero modes. The spacetime picture of the process
is a two-instanton process where $k$ fermion zero modes of the two
instantons are contracted against each other. In particular, in our
gauge theory example, the instanton $C_2$ induces a 4d effective
operator corresponding to a 4-fermion F-term, which then induces a
4-fermion interaction term on the effective action for the zero modes
of instanton $C_1$. Since the instanton $C_1$ has six fermion zero
modes, the lifting of four leaves only the two Goldstinos, so that
there is a non-trivial contribution to the superpotential.

To conclude, we would like to add yet another equivalent, but related,
viewpoint on the non-perturbative lifting of zero modes. The idea is
based on a generalization of the analysis in section 4.3 of
\cite{Beasley:2004ys}, which discussed the effects of a (perturbative)
superpotential mass term on instantons with additional fermion zero
modes. Consider an instanton $A$ with $k=2n$ fermion zero modes beyond
the two Goldstinos, and leading to a 4d higher F-term (\ref{bwop})
\begin{eqnarray}
\int \,d^4x\, d^2\theta\, {\cal O}_{w}\,
=\, \int \,d^4x\, d^2\theta\, \, w_{\ov i_1\ov j_1\ldots \ov i_n\ov j_n} (\Phi)\, {\cal O}^{\ov i_1\ov j_1} \ldots {\cal O}^{\ov i_n\ov j_n} 
\end{eqnarray}
with
\begin{eqnarray}
{\cal O}_{\ov i\ov j}\, =\, {\ov D} {\ov \Phi}^{\ov i}\, {\ov D} {\ov \Phi}^{\ov j}
\end{eqnarray}
The operator ${\cal O}_w$ is chiral (despite its appearance). In the
presence of an additional superpotential $W(\Phi)$, the supersymmetry
algebra is modified (since the fermion variations change,
$\delta\psi=F=-{\ov {\partial W/\partial \Phi}}$) and ${\cal O}_w$ is no longer
chiral. Still, since the instanton $A$ remains BPS, it should induce
an F-term. Indeed, in \cite{Beasley:2004ys} it was argued that (for
superpotential mass terms), there is a suitable deformation ${\tilde
  {\cal O}}_w$ of ${\cal O}_w$ which is chiral in the presence of the
superpotential. The instanton amplitude is now given by
\begin{eqnarray}
  \int \,d^4x\, d^2\theta\, {\tilde {\cal O}}_{w}\,
  =\, \int \,d^4x\, d^2\theta\, \, w_{\ov i_1\ov j_1\ldots \ov i_n\ov j_n} (\Phi)\, {\tilde {\cal O}}^{\ov i_1\ov j_1} \ldots {\tilde {\cal O}}^{\ov i_n\ov j_n} 
\end{eqnarray}
where, generalizing the result in \cite{Beasley:2004ys}, ${\tilde
  {\cal O}}_{\ov i\ov j}$ has schematically the structure
\begin{eqnarray}
  {\tilde {\cal O}}^{\ov i\ov j}\, =\, {\ov D} {\ov
    \Phi}^{\ov i}\, {\ov D} {\ov \Phi}^{\ov j}\, +\, W^{\ov i\ov j}
\label{supoinst}
\end{eqnarray}
Note that the total effect is that the instanton generate effective
vertices not only with $2n$ fermionic external legs, but also with
$2n-2p$ fermionic external legs (with $p$ taking several possible
values, depending on the detailed structure of $W$). The 4d
interpretation is that $2p$ fermionic legs have been soaked up by $p$
insertions of the superpotential interaction.

In fact, one is lead to suspect a further generalization of the above
argument. Consider the instanton $A$ in the presence, not of a 4d
superpotential term, but of a higher F-term (which could be of
perturbative or non-perturbative origin). Consider the latter to be of
the form
\begin{eqnarray}
  \delta S_{4d} \, =\, \int \, d^4x\, d^2\theta\, W_{\ov i_1\ov j_1\ldots \ov i_m\ov j_m} (\Phi)\, 
  {\ov D} {\ov \Phi}^{\ov i_1}\, {\ov D} {\ov \Phi}^{\ov j_1} \ldots 
  {\ov D} {\ov \Phi}^{\ov i_m}\, {\ov D} {\ov \Phi}^{\ov j_m}
\label{fterm}
\end{eqnarray}
Namely it leads to 4d interactions with $2m$ 4d fermions, and we
assume $m<n$. Although we do not have a precise argument based on the
supersymmetry algebra, we expect the amplitude of the instanton $A$ to
be modified in the presence of such term in the 4d action. Let us
define ${\tilde n}=n \text{ mod } m$ and $r=(n-{\tilde n})/m$, hence
$n=rm+{\tilde n}$. The instanton amplitude is expected to take the
schematic form
\begin{eqnarray}
  \int \,d^4x\, d^2\theta\, w_{ \{ \ov i_{1}\ov j_{1}\} \ldots
    \{ \ov i_{r}\ov j_{r}\}  \ov p_{\tilde n}}
  {\cal O}^{ \{ \ov i_{1}\ov j_{1}\} } \ldots
  {\cal O}^{\{ \ov i_{r}\ov j_{r}\} }\, 
  {\ov D} {\ov \Phi}^{\ov k_1}\, {\ov D} {\ov \Phi}^{\ov p_1} \ldots 
  {\ov D} {\ov \Phi}^{\ov k_{\tilde n}}\, {\ov D} {\ov \Phi}^{\ov p {\tilde n}}\,\nonumber
\end{eqnarray}
where $\{i_q,j_q\}$ denotes an $m$-plet of indices
$i_{q1},j_{q1}\ldots i_{qm}j_{qm}$, and
\begin{eqnarray}
  {\cal O}^{ \{ \ov i\ov j\} }\, =\, {\cal O}^{\ov
    i_{1} \ov j_{1}\ldots \ov i_{m} \ov j_{m}} \, =\, {\ov D} {\ov
    \Phi}^{\ov i_1}\, {\ov D} {\ov \Phi}^{\ov j_1} \ldots {\ov D} {\ov
    \Phi}^{\ov i_m}\, {\ov D} {\ov \Phi}^{\ov j_m} \, +\, W^{\ov
    i_1\ov j_1\ldots \ov i_m\ov j_m}
\label{fterminst}
\end{eqnarray}
The interpretation is that in the presence of the 4d F-term
(\ref{fterm}), the instanton with $2n$ fermion zero modes can generate
effective vertices with $2n-2m$ external fermionic legs, by having
sets of $2m$ fermionic legs soaked up by the F-term (\ref{fterm}).

The above discussion can be carried out to the situation where the
modification to the 4d action is induced by a second instanton $B$
with $2m$ fermion zero modes (which could be a gauge instanton or a
non-gauge D-brane instanton). In the spacetime picture, we would have
a multi-instanton process involving $A$ and $B$, in which some of the
fermionic external legs of the instanton $A$ are soaked up by the 4d
effective interaction induced by $B$. A simple example would be to
consider the instanton $B$ to have two fermion zero modes, so it
generates a superpotential, thus fitting into the situation leading to
(\ref{supoinst}). In fact, a particular case fitting within the
analysis in \cite{Beasley:2004ys} can be obtained by considering the
instanton $B$ to be a non-gauge D-brane instanton inducing a
superpotential mass term in the 4d action. Explicit examples of this
have been considered e.g. in
\cite{Bianchi:2007fx,Argurio:2007qk,Ibanez:2007tu}. Our example of
gauge theory instantons above corresponds to a more general situation
of the kind (\ref{fterminst}), with the instantons $A$, $B$ given by
the instantons $C_1$, $C_2$ (and $n=3$, $m=2$)

As a last remark, we expect processes with non-gauge instantons to
admit a similar interpretation. Thus the contribution to the
superpotential arising from the two-instanton process involving the
$U(1)$ and the $O(1)$ instantons can be regarded as the 4d effective
term induced by the $U(1)$ instanton in the presence of the additional
4d interaction induced by the $O(1)$ instanton.

\subsection{Adding semi-infinite D-branes}
\label{addingflav}

It is interesting to consider some simple modifications of the above
discussion in the presence of additional semi-infinite D6-branes
sticking out of the $\IC^*$ degenerations. From the field theory
viewpoint they correspond to the addition of extra flavours for some
of the gauge factors. From the viewpoint of the instantons, they lead
to additional fermion zero modes. In this section we consider a few
possibilities

In the above situation we have focused on a case where the
non-perturbative dynamics reduces to that of pure SYM. However, it is
straightforward to modify the setup to SQCD with $N_f$ flavors. It
suffices to introduce a stack of $N_f$ D6-branes wrapping the
non-compact 3-cycle obtained from a horizontal semi-infinite line
starting from the degeneration $a_2$ (this can be regarded as a limit
of infinite 3-cycle volume of a geometry with a second $b$-type
degeneration, located on the far right of the figure).  The above
argument goes through, and implies the continuity of the
non-perturbative superpotential across the line of marginal
stability. Notice that in the particular case of $N_f=N-1$ the
instantons under discussion are familiar gauge theory instantons.

Another straightforward addition of semi-infinite branes is to
consider adding $K$ D6-branes stretching from the $a_1$ degeneration
horizontally to the left infinity. Note that for the configuration in
Figure \ref{crossgauge}a, these D6-branes hit the $b$ degeneration, so
the configuration can be regarded as $K$ D6-branes stretching along
$(-\infty,b]$, $N+K$ on $[b,a_1]$ and $N$ on $[a_1,a_2]$. For the
configuration in Figure \ref{crossgauge}c, we have $N$ D6-branes on
$[b,a_2]$ and a disconnected set of $K$ D6-branes from left infinity
to $a_1$.

It is easy to carry out an analysis similar to the above to derive the
continuity of the superpotential. In the initial configuration, the
gauge factor $SU(N+K)$ has $N_f=N_c$ and thus a Beasley-Witten
instanton deforming its moduli space and forcing the gauge factor onto
the baryonic branch. The adjoint of the $SU(N)$ factor pairs up with
some of the mesons and becomes massive, so the left over pure SYM
theory develops a gaugino condensation superpotential. One recovers
the same result from the instanton contribution in the final
configuration Figure \ref{crossgauge}c (upon matching of scales along
the lines in Section \ref{introinst}).

\subsection{Gauge theory instantons and Seiberg duality}
\label{sec:seiberg-duality}

In this section we elaborate on an interesting point. It is a familiar
fact that the realization of Seiberg duality in terms of the D-brane
construction of gauge theories corresponds to a motion in moduli space
(in which D-branes typically break up and recombine)
\cite{Elitzur:1997fh,Ooguri:1997ih,Beasley:2001zp,Feng:2001bn,Cachazo:2001sg}
(see also \cite{Ito:1999xn,Berenstein:2002fi,Herzog:2004qw} for other
related approaches). Therefore they provide a large class of examples
of motion across lines of marginal stability in which the
non-perturbative superpotential is continuous.

A comment is in order here. From field theory experience we know that
Seiberg duality involves a non-trivial change of variables in the 4d
chiral multiplets. We also know that tree-level superpotentials are
crucial in matching properties of two Seiberg-dual theories. Both
properties are related to the following fact. Seiberg dualities in the
D-brane realization of field theories can be described as a motion
between two points $P$ and $Q$ in moduli space, at each of which we
have D-branes wrapped on cycles, whose sizes control the gauge
couplings and thus the strength of instanton effects. This motion
typically involves a region in moduli space larger than the radius of
convergence of the instanton expansion at either point. In other
words, the operation can also be described as a continuation past
infinite coupling, in the sense that they can be obtained by shrinking
a cycle $C$ on which 4d space filling branes wrap and growing a cycle
C' which is in the opposite homology class $[C']=-[C]$. The point $O$
where the cycle shrinks is strongly coupled from the viewpoint of the
original instanton at $P$, but a different weakly coupled description
is available at $Q$ (and vice versa). The change of description has
several effects, which we be taken into account implicitly in our
discussions below:

$\bullet$ It relates the strengths of the instantons as
$e^{-T}=(e^{-T'})^{-1}$, where $T$, $T'$ control the sizes of $C$,
$C'$. This underlies the fact that matching of scales in the Seiberg
duality encodes the continuity of the superpotential as a function of
the closed string moduli.

$\bullet$ It implies a non-trivial change of variables in the 4d
chiral multiplets, hence the comparison of the superpotentials at $P$
and $Q$ requires expressing the open string 4d multiplets in terms of
gauge invariant operators.

$\bullet$ It can map tree-level and non-perturbative superpotentials
to each other.  Thus the continuity applies to the full
superpotential.

The D-brane realization of Seiberg duality for large classes of field
theories thus provides a large class of examples of continuity of the
non-perturbative superpotential across lines of marginal stability
(with the appropriate change of variables for the charged matter
fields). We restrict to the description of this phenomenon with simple
examples, which are illustrative for this whole class.

\begin{figure}
\begin{center}
  
 \input{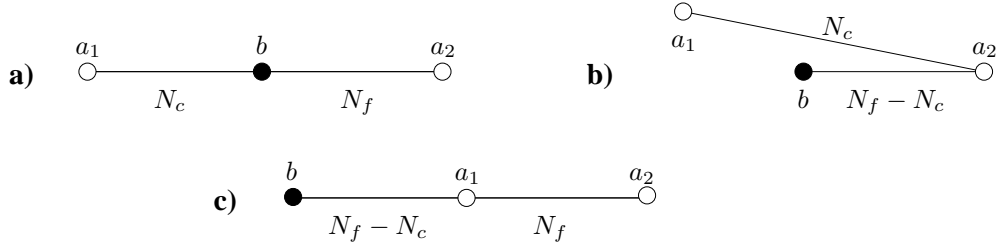}

  \caption{\small Realizing Seiberg duality in terms of D-branes: a)
    The electric configuration. b)~We move $a_1$ up a bit. The
    original branes are now nonaligned, so they recombine to minimize
    their tension. c)~Finally moving $a_1$ all the way to the middle
    position we get the magnetic dual theory.}
  \label{iia-emduality}
\end{center}
\end{figure}

Notice that it is easy to provide a D-brane realization of the
original Seiberg duality \cite{Seiberg:1994pq} using the above
geometries following \cite{Ooguri:1997ih}, as we review now, see
Figure \ref{iia-emduality}.

Consider a geometry with three aligned degenerations ordered as
$a_1,b,a_2$, and introduce $N_c$ D6-branes on $[a_1,b]$ and $N_f$ on
$[b,a_2]$, with $N_f\geq N_c$, Figure \ref{iia-emduality}a. This
describes the electric theory of $SU(N_c)$ SQCD with $N_f$ flavours,
with a gauged flavour group. Now move up the degeneration $a_1$. The
minimal energy configuration is obtained when $N_c$ D6-branes
recombine at $b$, so we have $N_c$ D6-branes on the tilted
\footnote{The tilting breaks supersymmetry in the intermediate steps
  of the argument; there are however simple modifications of the setup
  which allow to preserve supersymmetry throughout the process
  \cite{Ooguri:1997ih}. We skip their discussion since they will not
  be needed in our examples below.} segment $[a_1,a_2]$ and $N_f-N_c$
on $[b,a_2]$, Figure \ref{iia-emduality}b. Now move $a_1$ to the right
and bring it down between $b$ and $a_2$. The $N_c-N_f$ D6-branes on
$[b,a_2]$ split, so we are left with $N_f-N_c$ D6-branes on $[b,a_1]$
and $N_f$ on $[a_1,a_2]$. This describes the magnetic theory (again
with gauged flavour group). Note that the gauging of the flavour group
is just for the purposes of introducing configurations to be used
later; a realization of the pure Seiberg duality can be obtained
simply by sending the degeneration $a_2$ to right infinity.

Clearly the possibility of embedding Seiberg dualities in terms of
D-branes provides a huge class of examples of brane systems crossing
walls of marginal stability. The continuity of the non-perturbative
superpotential in these processes is automatically guaranteed by the
field theory argument for the matching of scales, as discussed
above. We will not delve into a more detailed discussion, and simply
discuss some particular examples related to systems in other Sections.

Let us focus on some particularly simple examples where the basic
splitting processes of the D6-branes are of the kind analyzed in the
previous Section. Consider the situation with $N$ D6-branes on
$[b,a_2]$ and no D6-branes on $[a_1,b]$. The $a_1$ degeneration has no
D6-branes attached, so moving it between the degenerations $b$, $a_2$
is exactly the inverse process of the one in Figure~\ref{crossgauge},
studied in Section~\ref{sqcd}.

For future convenience let us consider another example, now involving
semi-infinite D6-branes. Consider the initial configuration with
degenerations ordered as $a_1$, $b$, $a_2$ and introduce $N$ D6-branes
on $(-\infty,a_1)$, no D6-branes on $[a_1,b]$ and $K$ D6-branes on
$[b,a_2]$. As one moves the $a_1$ degeneration between $b$, $a_2$, it
drags the $K$ semi-infinite D6-branes, which end up split in the final
configuration. In the latter we have $K$ D6-branes on $(-\infty,b)$,
$N+K$ D6-branes on $(b,a_1)$, and $N$ D6-branes on $(a_1,a_2)$. The
splitting process of the semi-infinite D6-branes is exactly as in the
last system of Section \ref{addingflav}, where we showed the
continuity of the non-perturbative superpotential.

\medskip

\begin{figure}
\begin{center}
  
 \input{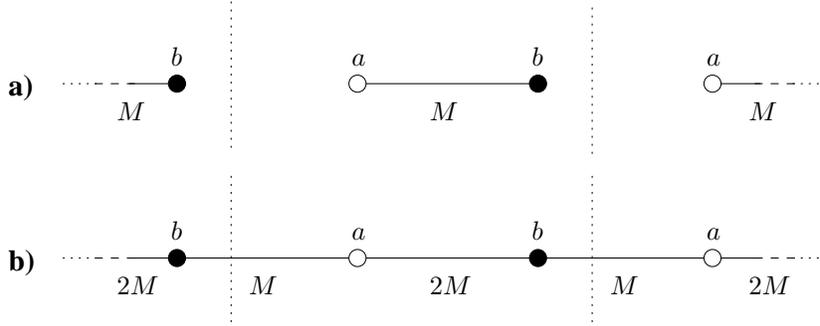}

  \caption{\small The periodic configuration dual to the conifold. The
    dotted vertical line denotes the period. a)~Final step of the
    cascade. b)~One step up in the cascade. We reach this point by
    moving all $a$ degenerations one cell to the left.}
  \label{iia-ks}
\end{center}
\end{figure}

As a final example based on the configuration in the previous
paragraph, let us consider a type IIA configuration mirror to D-branes
at the conifold, and (one of the steps of) the celebrated
Klebanov-Strassler duality cascade \cite{Klebanov:2000hb}. Following
\cite{Uranga:1998vf,Dasgupta:1998su}, a system of D-branes at a
conifold can be realized in terms of D4-branes suspended (along a
circle direction) between two rotated NS-branes. Equivalently, one can
use an infinite periodic array of rotated NS-branes with suspended
D4-branes. This systems can be mapped to one of our familiar double
$\IC^*$-fibration geometries by simply introducing a periodic array of
degenerations $\ldots,a,b,a,b,\ldots$, with D6-branes on the finite
segments, as shown in Figure \ref{iia-ks}. This is equivalent to (but
easier to visualize than) a double $\IC^*$ fibration over a cylinder,
with one degeneration of each type.

Consider the configuration on Figure \ref{iia-ks}a, with $M$ D6-branes
on the intervals of type $[a,b]$, and no D6-branes on those of type
$[b,a]$. This describes the theory at the end of the duality cascade,
and corresponds to $SU(M)$ SYM, with a non-perturbative superpotential
induced by a $1/M$-fractional instanton. Consider now the geometric
operation that takes us one step up the cascade. This corresponds to
moving the $a$-type degenerations once around the period, coming back
to its original position in the periodically identified geometry but
moving one period to the left in the covering space we are drawing. We
do this in the same way as above: moving up the $a$ singularity a bit,
taking it one cell to the left, and finally returning it to its
original vertical position. The resulting configuration is shown in
Figure \ref{iia-ks}b and contains $M$ D6-branes on the $[b,a]$
intervals and $2M$ D6-branes on the $[a,b]$ intervals. The geometric
process, and in particular the splitting of branes, is exactly as that
considered two paragraphs above, for $K=N\equiv M$. The continuity of
the superpotential is easily derived, by showing (using the instanton
interpretation of the field theory analysis in \cite{Gubser:2004qj})
that the Beasley-Witten instanton of the $SU(2M)$ theory (which has
$N_f=N_c$) deforms the moduli space of this theory and forces it into
the baryonic branch, while the $1/M$-fractional instanton on the left
over $SU(M)$ theory (with scale suitably computed by matching)
generates the superpotential.

\section{Exotic instantons becoming gauge instantons}
\label{sec:exotic-to-gauge}

In the previous Sections we have argued continuity of the
non-perturbative superpotential for gauge and non-gauge D-brane
instantons, in several examples. In this Section we would like to
consider a slightly more general situation where the nature of the
instanton changes in the process of reaching lines of marginal
stability. Namely a non-gauge D-brane instanton ends up as a gauge
D-brane instanton after some motion in moduli space.

A prototypical situation where this takes place is in duality cascades
\cite{Klebanov:2000hb} (see also
e.g. \cite{Franco:2004jz,Herzog:2004tr,Franco:2005fd,Brini:2006ej}) of
quiver gauge theories, in which one of the nodes of the quiver becomes
eventually empty of 4d space filling branes. D-brane instantons which
occupied this node change from gauge to non-gauge instantons in the
motion in moduli space associated to the cascade.  Since we are
interested in studying contributions to the superpotential, one would
need to consider cascades of orientifolded quiver gauge theories. In
fact, this kind of analysis has been carried out in
\cite{Aharony:2007pr} in one particular example, focusing on the
relevant part of the superpotential for the infrared theory. In
Section \ref{sec:cascade} we revisit the system in our language, and
recover that the full superpotential is well-behaved in the
process. Our analysis reproduces some pieces dropped in
\cite{Aharony:2007pr}, which are irrelevant in the infrared, but are
still part of the full superpotential of the theory.

Before revisiting the example of the duality cascade, let us consider
the simplest case where a non-gauge D-brane instanton becomes a gauge
theory effect.

\subsection{Dualizing the $O(1)$ instanton}
\label{sec:dualizing-O(1)}

\begin{figure}
\begin{center}
  
 \input{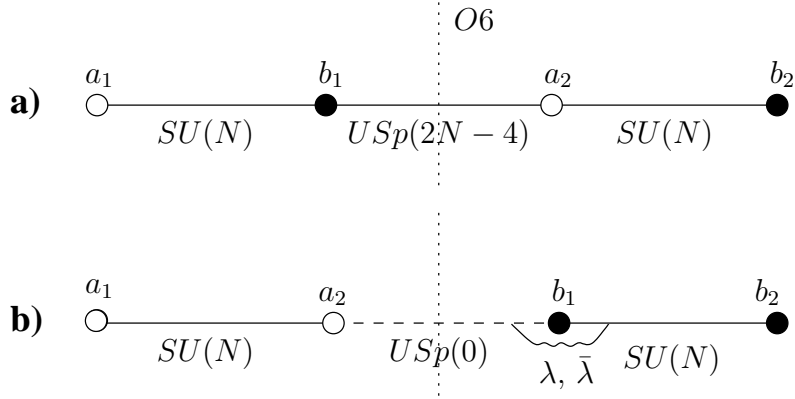}

  \caption{\small Prototypical example of a D-instanton effect being
    equivalent to a gauge theory effect via Seiberg duality. Figure a)
    shows the geometry leading to a $USp\times SU$ gauge theory upon
    wrapping D6-branes on the appropriate 3-cycles. The dotted line
    denotes the orientifold plane. Figure b) shows the configuration
    after the motion in moduli space corresponding to Seiberg
    duality. There are no D6-branes in the 3-cycle $[a_2,b_1]$, but an
    instanton (dashed line) wrapped on it can contribute to the
    superpotential. We have indicated the charged fermionic zero modes
    $\lambda$, $\bar \lambda$ between the D-brane instanton and the
    gauge D-brane.}
  \label{fig:gauge-D}
\end{center}
\end{figure}

Let us consider a geometry of the kind in Section \ref{geometries},
with an O6-plane, see Figure \ref{fig:gauge-D}. Let us wrap stack of
D6-branes on the different 3-cycles corresponding to the
configuration in Figure \ref{fig:gauge-D}a. The low energy dynamics of
this configuration is given a $SU(N)\times USp(2N-4)$ gauge theory,
with quarks $q\in (\fund_{SU},\fund_{USp})$, $\tilde q \in
(\antifund_{SU},\fund_{Usp})$ and superpotential
\begin{equation}
  W = q\tilde q q \tilde q
\end{equation}
Let us focus on the strong dynamics for the $USp$ theory
\footnote{There are additional non-perturbative effects from the
  $SU(N)$ factor, which can also be followed along the transition
  below, in analogy with our examples above (in fact, they map to
  2-instanton effects after the transition). We skip their discussion
  in order to emphasize the main point.}. As argued in
\cite{Intriligator:1995ne}, when the $USp$ node becomes strongly
coupled the theory has an effective description (corresponding to its
Seiberg dual) in which the $USp$ group confines completely, and the
fundamental degrees of freedom are the mesons:
\begin{equation}
  M_{\Yasymm}=q \cdot q\ ; \qquad
  M_{\bYasymm}=\tilde q \cdot \tilde q\ ; \qquad
  M_{Adj}=q\cdot \tilde q
\end{equation}
where we have expressed the mesons in terms of the electric fields
\footnote{We have omitted here the meson singlet under the $SU(N)$. In
  the stringy setup is will get a mass due to a coupling related by
  the $\cn=2$ susy to the one giving mass to the $U(1)$ gauge boson.},
and the dot denotes contraction in the $USp$ indices, which
antisymmetrizes the fields. The subindex denotes the representation of
the $SU(N)$ group under which the meson transforms. There is also a
superpotential implementing the classical constraint between the
mesons, which can be written as
\begin{equation}
  W_0 = \text{Pf} \begin{pmatrix}
    M_{\Yasymm} & M_{Adj} \\
    - M_{Adj} & M_{\bYasymm}
    \end{pmatrix}
\end{equation}
Adding the original superpotential in terms of the mesons we obtain
\begin{equation}
  W = W_0\, +\,  M_{\bYasymm} M_{\Yasymm}
\end{equation}
We can solve the equations of motion for the massive mesons
$M_{\bYasymm}$, $M_{\Yasymm}$ just by setting them to zero. The
resulting superpotential is then given by:
\begin{equation}
  \label{eq:W-gauge-theory}
  W = \text{Pf} \begin{pmatrix}
    0 & M_{Adj} \\
    -M_{Adj} & 0
    \end{pmatrix} = \det M_{Adj}.
\end{equation}

We can now perform a brane motion taking the configuration to that in
Fig.~\ref{fig:gauge-D}b, where there is no brane stretching on the
3-cycle $[a_2,b_1]$.  This result takes into account the brane
creation effects due to the presence of the orientifold planes, as
discussed in \cite{Ooguri:1997ih}. Despite the non-trivial change in
the brane configuration, the superpotential is continuous. Namely the
above superpotential is still generated, but now via an exotic $O(1)$
instanton on $[a_2,b_1]$ which can contribute \footnote{Recall that
  the orientifold projection acts oppositely on 4d space filling
  D6-branes and D2-brane instantons, so an orientifold giving a $USp$
  gauge group will give a $O(1)$ D-instanton. This works in the same
  way as for the perhaps more familiar D5-D9 system.}. The calculation
in this case is simple. In Figure \ref{fig:gauge-D}b, the theory on
the $SU(N)$ brane is locally $\cn=2$, in particular it has an adjoint,
which we identify with the adjoint meson of the gauge analysis (in
both cases it parametrizes sliding the D6-branes along the two
$b$-type degenerations, and their images along the $a$-type ones). The
zero modes $\lambda$, $\bar\lambda$ between the D2-brane instanton and
the $SU(N)$ brane couple to this adjoint via a term
\begin{equation}
  S = \ldots + \lambda M_{Adj} \bar \lambda
\end{equation}
in the instanton action (this has the same origin as the usual
coupling between the adjoint and the flavors in $\cn=2$ theories).
Integrating over the fermionic zero modes gives us the determinant
operator we found in equation \ref{eq:W-gauge-theory}. We thus recover
the same kind of superpotential, with an exponential dependence on the
closed string modulus associated to the 3-cycle defined by the
degenerations $b_1$ and $a_2$. Thus the result is continuous across
the motion in moduli space, in which gauge and non-gauge instantons
turn into each other \footnote{\label{noncompact} One may try to
  discuss a similar configuration without the external degenerations
  and with semi-infinite branes. In this case the computation
  \ref{fig:gauge-D}a gives a non-zero superpotential, while in
  \ref{fig:gauge-D}b there are no dynamical mesons to help saturate
  the fermion zero modes of the instanton, hence there is no
  superpotential. The mismatch is related to the non-compact D-branes
  in the configuration (see comment at the beginning of Section
  \ref{geometries}). Upon ``compactification'' by adding the external
  degenerations at a finite distance, one recovers the above full
  agreement.}.

\subsection{A duality cascade example}
\label{sec:cascade}

Let us proceed to the more complicated case of the duality cascade
studied in \cite{Aharony:2007pr}, and show the continuity of the
non-perturbative superpotential along a complicated chain of Seiberg
dualities.

\begin{figure}
\begin{center}
  
 \input{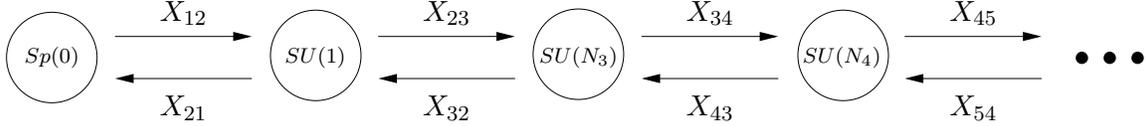}

  \caption{\small Relevant nodes of the quiver theory for the
    orbifolded conifold. We have indicated the ranks at the bottom of
    the cascade. There can be more $SU(N)$ nodes to the right, ending
    with another $USp(N)$ group. $X_{ij}$ denotes the bifundamental
    from node $i$ to node $j$.}
  \label{AK-bottom}
\end{center}
\end{figure}

The theory under consideration is given by the quiver in Figure
\ref{AK-bottom}, with gauge group at the bottom of the cascade given
by $USp(0)\times SU(1)\times SU(N_3)\times\ldots$ with $N_3,\ldots$
arbitrary. The superpotential is given by:
\begin{equation}
  W = \sum_{i=1}^{N_{factors}} (-1)^i X_{i,i+1} X_{i+1,i+2} X_{i+2,i+1} X_{i+1,i}.
  \label{eq:W-conifold}
\end{equation}

This theory can be easily realized in string theory by modding out an
orbifold of the conifold \cite{Uranga:1998vf} by a suitable orientifold action
\cite{Franco:2007ii}. In terms of the geometries in Section
\ref{geometries}, we can consider a periodic array of degenerations
$a_1,b_1,a_2,b_2,a_3,b_3,a_4,b_4$ and introducing an orientifold
quotient $\Omega R(-1)^{F_L}$, with $R$ given by (\ref{orient}).

In terms of the geometrical setup, the cascade of Seiberg dualities
simply amounts to a motion in moduli space, generalizing that
discussed above. In this situation there are also some brane creation
effects due to the presence of the orientifold planes, as discussed in
\cite{Ooguri:1997ih}. The configuration one step up in the cascade is
given by the same quiver but with different ranks:
\begin{equation}
  USp(2N_2-4) \times SU(N_2) \times SU(N_4-1)\times SU(N_4)\times
  \ldots
\end{equation}
In particular all nodes are occupied so there are no non-gauge
instantons at this level.

The detailed gauge theory analysis of \cite{Aharony:2007pr} for the
initial configuration shows that the nonperturbative superpotential of
the initial gauge theory configuration can be described in terms of
the fields at the end of the cascade as
\begin{equation}
  W_{np}^{bottom} = X_{23} X_{32} + \det(X_{34} X_{43}) + X_{23}
  X_{34} X_{43} X_{32} + \ldots
\label{aksupo}
\end{equation}
where we have omitted some quartic terms of the same form as those in
(\ref{eq:W-conifold}), which are tree level from the point of view of
$g_s$, and thus not particularly interesting here. We rather us focus
on the first two terms, which are nonperturbative. Our aim is to
recover them by studying possible D-brane instantons in the final
configuration. Note that the determinant term was dropped as
irrelevant in \cite{Aharony:2007pr}, since they were just interested
in the infrared behaviour of the theory. We are interested in the
continuity of the full superpotential so we should keep it, since it
carries an implicit dependence on the closed string modulus
controlling the corresponding cycle. For completeness, let us
reproduce here a sketch of the gauge theory analysis done in
\cite{Aharony:2007pr}:

\subsubsection{The gauge analysis}

We will assume for simplicity a hierarchy of scales given by
\begin{equation}
  \Lambda_1 \gg \Lambda_3 \gg \ldots \gg \Lambda_2 \gg \Lambda_4 \gg
  \ldots
\end{equation}
We will choose the ranks in such a way that the bottom of the cascade
is described by the quiver in Figure~\ref{AK-bottom}. This can be
achieved by choosing the following ranks:
\begin{equation}
  N_1 = 2N_2 - 4 \quad ; \quad N_3 = N_4-1.
\end{equation}
Due to the hierarchy of scales we have chosen, the first node to
become strongly coupled is the $USp$ one. This goes just as in Section
\ref{sec:dualizing-O(1)}, and we end up with a $USp(0)$ group, some
mesons $M_{Adj}$ charged in the adjoint of $SU(N_2)$, and a
nonperturbative superpotential:
\begin{equation}
  W_{np} = \det M_{Adj}
\end{equation}

Now we have to dualize the $SU(N_3)$ node. We have $N_f=N_2+N_4$, so
the dual description is in terms of a $SU(N_f-N_c=N_2+1)$ gauge group,
and the dual quarks and mesons. The mesons get a mass due to the
quartic terms in the superpotential \ref{eq:W-conifold}, and they can
be integrated out. Also, there is a mass coupling coming from the
superpotential between $M_{Adj}$ and the meson $M_2^{(3)}$ of
$SU(N_3)$ charged under the adjoint of $SU(N_2)$. The relevant term in
the superpotential looks like:
\begin{equation}
  W = \ldots + \det M_{Adj} + M_{Adj} M_2^{(3)} + M_2^{(3)} q_{23}q_{32}
\end{equation}
with $q$ the dual quarks. Integrating the mesons out, we end up with a
superpotential:
\begin{equation}
  W = \ldots + \det q_{23} q_{32} + q_{23} q_{34} q_{43} q_{32}
\label{eq:det-qq}
\end{equation}
where we have included a piece of the quartic superpotential that will
play a role in a moment.

Going down in energy, eventually the $SU(N_2)$ node will become
strongly coupled. It has $N_f=N_2+1$, coming just from the third node,
so the gauge group confines completely (let us call the resulting node
``$SU(1)$'', as in the stringy picture of the duality there is a
single brane remaining). The description is in terms of mesons
$M_3^{(2)}$ in the adjoint of the third node and baryons $B_3$,
$\tilde B_3$ in the fundamental and antifundamental. There is a
superpotential given by:
\begin{equation}
  W = \ldots + B_3 M_3^{(2)} \tilde B_3 - \det M_3^{(2)}
\end{equation}
When the second node confines the $q_{23}$, $q_{32}$ quarks get
confined into baryons and mesons. In particular, the superpotential
\ref{eq:det-qq} can be expressed as:
\begin{equation}
  W = \ldots + B_3 \tilde B_3 + M_3^{(2)} q_{34}q_{43}
  \label{eq:supo-node2}
\end{equation}

The last step in the chain of dualities, as far as the first three
nodes are concerned, comes from dualizing the fourth node. This is
important for our discussion as it gives a mass to $M_3^{(2)}$ via the
dual of the last coupling in eq. \ref{eq:supo-node2}. After dualizing
node 4, we end up with a superpotential:
\begin{equation}
  W = \ldots + B_3 M_3^{(2)} \tilde B_3 - \det M_3^{(2)} + B_3 \tilde
  B_3 + M_3^{(2)} M_3^{(4)} - M_3^{(4)} X_{34}X_{43}
\end{equation}
where we have denoted as $X_{34}$, $X_{43}$ the dual quarks of node 4
charged under node 3. We see that the mesons of node 3 get massive as
expected. Integrating them out, one gets:
\begin{equation}
  W = \ldots + B_3 X_{34}X_{43}\tilde B_3 -\det X_{34}X_{43} + B_3
  \tilde B_3
\end{equation}
which is the same as the one in (\ref{aksupo}) up to a relabeling of
the baryons as $X_{23}$, $X_{32}$.

\subsubsection{D-instanton effects at the bottom of the cascade}

Let us now consider the final configuration, where the 4d space
filling D6-brane configuration gives rise to a structure $USp(0)\times
SU(1)\times SU(N_3)\times\ldots$ with $N_3,\ldots$. There are two
instantons which can contribute to the superpotential. There is a
non-gauge D-brane instanton arising from the cycle corresponding to
the node of the quiver with no 4d space filling branes. As argued in
\cite{Aharony:2007pr} and we now review, it leads to the first mass
terms in (\ref{aksupo}). The instanton has $O(1)$ symmetry and has two
neutral fermion zero modes. In addition it has two fermion zero modes
$\alpha$ and $\beta$ from the open strings going from the D-instanton
to the $SU(1)$ brane. The instanton action contains a coupling of the
form $\alpha X_{23} X_{32} \beta$, arising from the same disk
instantons that produce the terms in
(\ref{eq:W-conifold}). Integrating over these fermionic zero modes, we
get a mass contribution to the superpotential:
\begin{align}
  W = & \ldots + \int\!d\alpha d\beta \ \alpha X_{23} X_{32}
  \beta \notag\\
   = & \ldots + X_{23} X_{32}.
\end{align}

There is another D-brane instanton which contributes to the
non-perturbative superpotential, and which involves a somewhat novel
effect. It corresponds to a D-brane instanton on the node with 4d
group ``$SU(1)$''. This instanton does not have a proper gauge theory
interpretation, but still it shares some common features with gauge
instantons. Namely, since it is a $U(1)$ instanton, not mapped to
itself by the orientifold action, it has four fermion zero modes. The
two Goldstinos of $\cn=1$ supersymmetry remain, while the two
accidental $\cn=2$ Goldstinos have non-trivial couplings with the
bosonic and fermionic zero modes in the sector of open strings between
the instanton and the $SU(1)$-brane. For gauge D-brane instantons,
integration over these zero modes imposes the fermionic ADHM
constraints \cite{Billo:2002hm}, and reproduces the correct measure on
instanton (super)moduli space. In the present setup, we lack an
appropriate gauge theory interpretation for the coupling, but its
effect of leading to the saturation of the additional fermion zero
modes remains. We are therefore left with the two Goldstinos
$\theta^{\alpha}$ needed for contributing to the superpotential.  We
still need to saturate the charged zero modes going from the
D-instanton to the $SU(N_3)$ group, there are $2N_c$ of these, $N_c$
of each chirality. Let us call them $\lambda_{23}$ and
$\lambda_{32}$. They can be saturated via the same kind of quartic
coupling $\lambda_{23} Y_{34} Y_{43} \lambda_{32}$ as above. Expanding
the instanton action we get a contribution to the superpotential:
\begin{align}
  W = & \ldots + \int\![d\lambda_{23}][d\lambda_{32}]\ \exp\left(\lambda_{23}
  Y_{34} Y_{43} \lambda_{32}\right) \notag\\
  \simeq & \ldots + \epsilon^{i_1\ldots i_{N_3}}\epsilon^{k_1\ldots k_{N_3}}
  (Y_{34}Y_{43})_{i_1,k_1} \cdot \ldots \cdot
  (Y_{34}Y_{43})_{i_{N_3},k_{N_3}}\notag \\
  \simeq & \ldots + \det(Y_{34}Y_{43}).
\end{align}
which correctly reproduces the second term in the nonperturbative
superpotential \ref{aksupo}.

We see that there is a beautiful agreement between both
computations. Clearly, there are plenty of other systems where the
agreement between the superpotential up in the cascade and at the
lower steps can be checked. We leave this analysis for the interested
reader.

\section{Topology changing transitions in F-theory}
\label{sec:F-theory}

In this section we comment on an intriguing implication of the
continuity of the non-perturbative superpotential, when considering
the F-theory viewpoint on non-perturbative effects on systems of
D7-branes near lines of marginal stability. The process of D7-branes
splitting/recombining corresponds to a topology changing transition in
F/M-theory, along the lines of \cite{Uranga:2002ag}. Our results
therefore imply a non-trivial relation between the non-perturbative
superpotentials on topologically different Calabi-Yau fourfolds.

We restrict to a simple local analysis of such D7-brane system, and of
its F-theory lift.  Consider the type IIB D7-brane realization of the
D-brane configuration studied in Section \ref{sqcd}. There are two
stack of D7-branes wrapped on two holomorphic 4-cycles $C_1$ and
$C_2$, intersecting over a complex curve $\Sigma$. It is possible to
consider concrete examples of Calabi-Yau threefolds and 4-cycles with
$h_{2,0}(C_1)=1$, $h_{2,0}(C_2)=0$, which would fit our example, but
it is not necessary to illustrate the main point. In fact, the basic
idea is already present in a local model in a neighborhood of a point
$P$ in $\Sigma$. Using local complex coordinates $z,w,u$ we have
D7-branes on $C_1$, described locally by $w=0$ (and $z,u$ arbitrary)
and D7-branes on $C_2$, described locally by $z=0$ (and $w,u$
arbitrary). The curve $\Sigma$ is locally parametrized by $u$. In this
local analysis, the direction $u$ is an spectator and we can ignore it
in the following (although it can lead to global obstructions in the
compact model). Thus we have a system of D7-branes wrapped on the
locus $zw=0$.

The F-theory lift of this configuration is described by an elliptic
fibration over the threefold, with degenerate fibers (due to pinching
of a 1-cycle) over the 4-cycle wrapped by the D7-branes. We can also
work locally near the pinching of the elliptic fiber, and describe the
geometry as a $\IC^*$ fibration. For $n$, $m$ D7-branes on the two
different 4-cycles, the local description of the fourfold is thus
given by the spectator direction $u$ times the manifold
\begin{eqnarray}
xy=z^nw^m
\end{eqnarray}
This kind of geometries were introduced in \cite{Uranga:1998vf}. Let
us focus on the simplest representative, $n=m=1$, the conifold. In
fact, the configuration corresponds to the resolved conifold, with the
2-cycle described as follows. The fiber on top of the intersection
locus $z=w=0$ on the base degenerates into two 2-spheres touching at
two points. The class of the 2-cycle corresponds to one of these
2-spheres (while the sum is the class of the fiber). For intersecting
D7-branes, the F/M-theory lift corresponds to the limit of vanishing
2-cycle (and no background 2-form potential can be turned on). We are
thus at the singular conifold limit, in which there are massless
states \cite{Strominger:1995cz} (arising from wrapped M2-branes in the
M-theory picture). These are nothing but the open strings degrees of
freedom between the two D7-brane stacks.

Consider now the D7-brane system away from the line of marginal
stability. The D7-branes recombine into a single smooth one, wrapped
on a 4-cycle which is a deformation of the above, namely
$zw=\epsilon$. In the local model, $\epsilon$ corresponds to a
modulus, a flat direction for the fields arising at the intersection
of the D7-branes. In the global model the flat direction is obstructed
by a D-term condition, and the value of $\epsilon$ is fixed by the
closed string modulus moving us away from marginal stability.  The
F-theory lift of this configuration corresponds to the geometry
\begin{eqnarray}
xy = zw -\epsilon
\end{eqnarray}
This describes the deformed conifold. This is expected, since the
massless charged states have acquired a vev, thus triggering a
topology changing transition \cite{Greene:1995hu}. The behaviour of
the arbitrary $n=m$ case is similar, using the deformation
$xy=(zw-\epsilon)^n$.

The local analysis shows that the crossing of a line of marginal
stability corresponds to a topology change in the F/M-theory
fourfold. The continuity of the non-perturbative superpotential in
this case implies a non-trivial matching between topologically
different spaces.

It would be interesting to have a more microscopic derivation of this
result. We conclude by mentioning a few key points to this aim. The
relevant instanton in the IIB picture is a D3-brane wrapped on the
4-cycle which splits at the line of marginal stability. As emphasized,
the continuity of the process requires a non-trivial contribution from
a 2-instanton process in the intersecting D7-brane configuration. In
the F/M-theory lift, the effect arises from an M5-brane instanton
wrapping a 6-cycle which splits, and there should exist a non-trivial
contribution to the superpotential arising from a 2-instanton process
involving the two M5-brane instantons wrapped on the two components of
the split 6-cycle. Thus our analysis of superpotentials from
multi-instantons should apply to M5-brane instantons on M-theory on CY
fourfolds. Clearly this goes beyond the analysis in
\cite{Witten:1996bn}, since one would require a suitable
generalization to M5-brane instantons on singular 6-cycles. In this
respect, notice that one can rephrase the multi-instanton process as a
non-perturbative lifting of zero modes of one M5-brane (A) by the
effects of a second M5-brane (B). This is not inconsistent with the
arguments in \cite{Witten:1996bn}, which were based on counting
fermion zero modes chiral with respect to the $U(1)$ symmetry acting
on the normal directions transverse to the M5-brane A. Indeed the
second M5-brane B can induce couplings which violate this $U(1)$
(which acts on directions along the volume of the M-brane B). Thus the
non-perturbative lifting mechanism is powerful enough to allow the
appearance of contributions from instantons which violate the
celebrated arithmetic genus condition in \cite{Witten:1996bn}.
Concrete examples of this are provided by suitable F/M-theory versions
of the type II models studied in this paper.

\section{Conclusions and outlook}
\label{sec:conclusions}

In this paper we have studied the microscopic mechanisms via which
D-brane instanton computations lead to non-perturbative superpotential
continuous across moduli space. This understanding has revealed
interesting surprises, including the interesting role of
multi-instanton contributions to the superpotential, and its
interpretation as non-perturbative lifting of fermion zero modes.

These results go in the direction that D-brane instanton effects are
subtler, and more abundant, than hitherto considered. It would be
interesting to revisit some of the models considered in the literature
and look for additional contributing instanton processes of the kind
we have introduced.

The computation of multi-instanton processes to the superpotential are
involved, and require the precise knowledge of the zero mode
interactions. It would be interesting to use the continuity of the
non-perturbative superpotential to systematize or short-cut such
computations. For instance, consider a set of BPS instantons $\{
C_i\}$ at a point $P$ in moduli space. If these instantons can form an
irreducible bound state $C$ somewhere else in moduli space (at a point
$Q$), and if $C$ has only two fermion zero modes, then in the theory
at $P$ there is a non-trivial multi-instanton process involving the
instantons $\{ C_i\}$. Similarly, if the instantons form a bound
state, but it has more than two fermion zero modes, the corresponding
superpotential (at $Q$ and hence at $P$) vanishes. This seemingly
innocent statement is in fact very powerful. For instance it may be
feasible to systematically construct instantons contributing to the
superpotential at some tractable point in moduli space, and translate
the corresponding instanton processes to the corresponding (possibly
multi-)instanton processes in other regions. For instance, BPS
instantons on type IIB models may be constructed as stable holomorphic
gauge bundles in the large volume regime. Such contributing instantons
could subsequently be translated into multi-instanton processes at
other interesting points like orbifold limits or Gepner points.

The non-perturbative superpotential is an interesting quantity which
is well-behaved all over moduli space, in a non-trivial way. It would
be interesting to gain a deeper understanding of the microphysics
underlying this result in general, beyond the concrete examples we
have analyzed. We expect further insights from more powerful
approaches, for instance using the category of holomorphic D-branes,
which is another interesting object with universal properties over
moduli space.  This category does not include the information about
the stability conditions on D-branes, namely on the D-term
contributions to the world-volume action. However our results suggest
that the full superpotential is rather insensitive to the stability
properties of individual BPS instantons: as soon as an instanton
become unstable and decays into sub-objects, the latter can
reconstruct the same contribution via a multi-instanton process.

We expect our results to shed light on the physics of non-perturbative
superpotentials in string theory, both from the viewpoint of its
formal properties, and for physical applications in concrete examples.

\vspace*{1cm}

{\bf Acknowledgments}\\
We thank C. Bachas, E. Kiritsis, L. Iba\~nez and D. Tong for useful
discussions. A.M.U. thanks M. Gonz\'alez for encouragement and
support. I.G.-E. thanks the CERN Theory division for hospitality, and
N. Hasegawa for kind support. This work has been supported by the
European Commission under RTN European Programs MRTN-CT-2004-503369,
MRTN-CT-2004-005105, by the CICYT (Spain) and the Comunidad de Madrid
under project HEPHACOS P-ESP-00346. The work of I.G.-E. is financed by
the Gobierno Vasco PhD fellowship program.

\appendix

\section{No splitting $O(1)\to O(1)\times O(1)$}
\label{nosplit}

In the main text we have described examples where an $O(1)$ instanton
decays into a set of two instantons of $U(1)\times O(1)$, or a $U(1)$
instanton and its orientifold image. In this appendix we show that it
is not possible to have one $O(1)$ instanton decay into two $O(1)$
instantons (or more generally, that an instanton mapped to itself
under the orientifold action cannot decay into two instantons
invariant under the orientifold action). The argument is general, and
applies to any BPS instanton, contributing to the superpotential or to
higher-fermion F-terms.

The general argument that forbids such instantons from
crossing such marginal stability lines goes as follows. In order
to contribute to F-terms, the instanton must be BPS, so the
cycle it wraps must calibrated by $e^{i\varphi}\Omega$ with some
constant phase $\varphi$:
\begin{equation}
  \arg \Omega|_{\Xi}=\pi \varphi.
\end{equation}
The phase $\varphi$ of the calibration determines what is called in
the $\Pi$-stability literature the {\em grading} of the brane, we will
adopt this terminology here. This grading determines the supersymmetry
conserved by the brane, and also the mass of the lightest bosonic
string mode between the decay products $\Xi_1$ and $\Xi_2$ at the
marginal stability line:
\begin{equation}
  m^2 = \frac{1}{2}(\varphi_1-\varphi_2).
\end{equation}

When the gradings $\varphi_1$ and $\varphi_2$ of the decay products
coincide $\Xi_1$ and $\Xi_2$ are mutually supersymmetric, the boson is
massless, and we are on the marginal stability wall. Once we move
slightly off the wall, the gradings will become different and bound
state formation becomes possible. Whether we have a bound state or a
stable superposition of two mutually non-supersymmetric branes depends
on the sign of the boson mass: on one side of the wall it will become
tachyonic, triggering bound state formation, while in the other side
of the wall it will be massive, and the superposition of $\Xi_1$ and
$\Xi_2$ is stable.

Here the main point of interest for our discussion is that the
$\Omega\rightarrow\overline\Omega$ action of the orientifold acts on
these gradings as $\varphi \rightarrow -\varphi$, so invariant
instantons must have integer grading. This obstructs the decay of the
$O(1)$ instanton into two $O(1)$ factors: there is no question of
continuity of the nonperturbative superpotential since the gradings of
the branes are frozen by the orientifold.

It is possible to argue the same thing in a slightly different
way. Imagine D6 branes wrapping the same cycles in the internal space
as the instantons. In this case the process of brane recombination is
typically seen as a Higgsing, triggered by a Fayet-Iliopoulos
term. The interpretation of the discussion above in terms of the gauge
theory living on the brane is that due to the orientifold, the gauge
group on the brane gets reduced from $U(1)$ to $O(1)$, and the
Fayet-Iliopoulos is projected out. There is no continuous way of
Higgsing $O(1)\times O(1)$ to $O(1)$.

\begin{figure}
\begin{center}
  
 \input{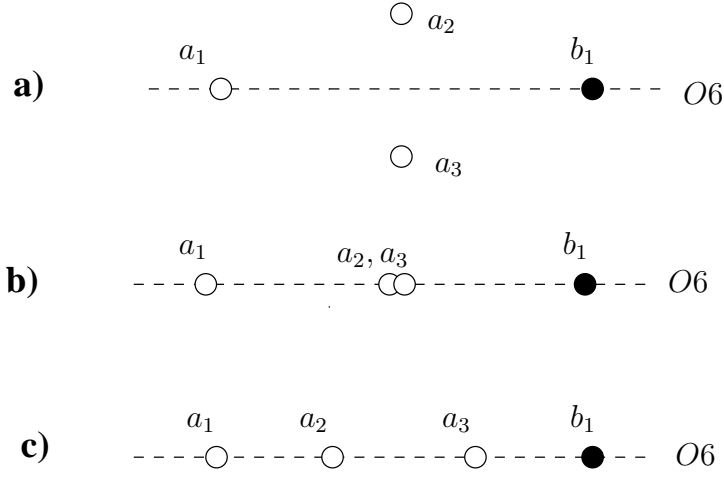}

  \caption{\small Splitting an instanton invariant under the
    orientifold action into two invariant instantons via a process
    involving a singular configuration.}
  \label{exotictrans}
\end{center}
\end{figure}

We conclude by pointing out that our argument above does not exclude
other more exotic possibilities to split an instanton invariant under
the orientifold action into two instantons invariant under the
orientifold action. In fact, there is a simple example of such
transition, which is related to those in \cite{Brunner:1998jr}, as we
now describe. Consider a geometry of the kind considered in Section
\ref{geometries}, as shown in Figure \ref{exotictrans}. The
configuration includes an orientifold plane associated to the action
$\Omega R(-1)^{F_L}$, with $R$ given by
\begin{eqnarray}
z\to {\ov z} \quad ; \quad (x,y) \to (\ov y,\ov x) \quad ; \quad (x'y') \to (\ov y',\ov x')
\label{orientfour}
\end{eqnarray}
or 
\begin{eqnarray}
z\to {\ov z} \quad ; \quad (x,y) \to (\ov x,\ov y) \quad ; \quad (x'y') \to (\ov x',\ov y')
\label{orienteight}
\end{eqnarray}
The two choices lead to orientifold planes whose projection on the
$z$-plane is the horizontal axis. They act differently on the $\IC^*$
fibers, and lead to slightly different structures for the orientifold
projection \footnote{The two choices correspond in the HW dual to
  introducing O4- or O8-planes, respectively.}. Namely the O6-plane
defined by (\ref{orientfour}) is split when it encounters a $\IC^*$
degeneration, and it changes from O6$^+$ to O6$^-$ (and vice
versa). The O6-plane defined by (\ref{orienteight}) is not split and
has a fixed RR charge. This distinction will not be relevant for us,
and for concreteness we focus on an orientifold of the kind
(\ref{orienteight}), and choose the orientifold to lead to $O(1)$
symmetries for D2-branes instantons.

Consider the transition shown in Figure \ref{exotictrans}. We consider
the fate of the $O(1)$ instanton arising from a D2-brane on the
3-cycle $[a_1,b_1]$. As the two degenerations $a_2$, $a_3$ approach
the orientifold plane (in a way consistent with the orientifold
action), we reach a singular configuration, Figure \ref{exotictrans}b,
where the $O(1)$ instanton is split into two $O(1)$ instantons. At
this point a new branch emerges, where $a_2$, $a_3$ can separate along
the horizontal axis, and the original instanton is split into three
$O(1)$ instantons (for the orientifold action (\ref{orientfour}), the
middle 3-cycle would lead to an $USp$ instantons). It is thus possible
to split $O(1)$ instantons by a physical process, but which is in fact
unrelated to (and more exotic than) lines of marginal stability.
Indeed, notice that the transitions is not triggered by a
Fayet-Iliopoulos parameter.

In fact, it is questionable that the transition has a simple
description from the viewpoint of the instanton world-volume. Notice
that the transition involves passing through a singular configuration,
on which the 2-sphere of the 3-cycle $[a_2,a_3]$ shrinks to zero
size. This is an orientifold quotient of the singular CFT point of the
$A_1$ geometry, where the theory (at least before orientifolding)
develops enhanced gauge symmetry, with additional massless gauge
bosons arising from wrapped D2-branes. It is unlikely that the
transition point admits a standard description from the viewpoint of
the instanton world-volume. We thus expect that the non-perturbative
superpotential can be discontinuous across this kind of
transition. Indeed, in \cite{Brunner:1998jr} similar transitions lead
to discontinuous phenomena, like chirality changing phase transitions,
not compatible with a local field theory description. It would be
interesting to investigate these transitions in more detail, perhaps
along the lines in \cite{Hori:2005bk}.

\end{document}